\documentclass[aps,twocolumn,amsmath,amsfonts,nofootinbib,preprintnumbers,
  superscriptaddress,eqsecnum,secnumarabic]{revtex4-1}
\usepackage{graphicx,slashed,hyperref}
\usepackage[utf8]{inputenc}
\hypersetup{
   bookmarks=true,         
   unicode=true,          
   pdftoolbar=true,        
   pdfmenubar=true,        
   pdffitwindow=false,     
   pdfstartview={FitH},    
   pdftitle={My title},    
   pdfauthor={Author},     
   pdfsubject={Subject},   
   pdfcreator={Creator},   
   pdfproducer={Producer}, 
   pdfkeywords={keyword1} {key2} {key3}, 
   pdfnewwindow=true,      
   colorlinks=true,       
   linkcolor=blue,          
   citecolor=magenta,        
   filecolor=magenta,      
   urlcolor=cyan           
}
\usepackage{amsmath}
\usepackage{amssymb}

\newcommand{\be}{\begin{equation}}
\newcommand{\ee}{\end{equation}}
\def\bsp#1\esp{\begin{split}#1\end{split}}

\newcommand{\ma}{{\sc Mad\-A\-na\-ly\-sis}~5}
\newcommand{\ms}{{\sc Mad\-Spin}}
\newcommand{\mw}{{\sc Mad\-Width}}
\newcommand{\fj}{{\sc Fast\-Jet}}

\newcommand{\del}{{\sc Del\-phes}}

\begin{document}
\title{Top-philic Vector-Like Portal to Scalar Dark Matter}

\author{Stefano Colucci}
\email{colucci@th.physik.uni--bonn.de}
\affiliation{Physikalisches Institut der Universit\"at Bonn, Bethe Center
  for Theoretical Physics, \\ Nu{\ss}allee 12, 53115 Bonn, Germany}

\author{Benjamin~Fuks}
\email{fuks@lpthe.jussieu.fr}
\affiliation{Sorbonne Universit\'e, CNRS, Laboratoire de Physique Th\'eorique et
  Hautes \'Energies, LPTHE, F-75005 Paris, France}
\affiliation{Institut Universitaire de France, 103 boulevard Saint-Michel,
  75005 Paris, France}

\author{Federica~Giacchino}
\email{federica.giacchino@ulb.ac.be}
\affiliation{Service de Physique Th\'eorique, CP225, Universit\'e Libre de Bruxelles, Bld du Triomphe, 1050 Brussels, Belgium}

\author{Laura~Lopez~Honorez}
\email{llopezho@ulb.ac.be}
\affiliation{Service de Physique Th\'eorique, CP225, Universit\'e Libre de Bruxelles, Bld du Triomphe, 1050 Brussels, Belgium}
\affiliation{Vrije Universiteit Brussel and The International Solvay Institutes,
Pleinlaan 2, 1050 Brussels, Belgium.}

\author{Michel~H.G.~Tytgat}
\email{mtytgat@ulb.ac.be}
\affiliation{Service de Physique Th\'eorique, CP225, Universit\'e Libre de Bruxelles, Bld du Triomphe, 1050 Brussels, Belgium}

\author{J\'{e}r\^{o}me~Vandecasteele}
\email{jvdecast@ulb.ac.be}
\affiliation{Service de Physique Th\'eorique, CP225, Universit\'e Libre de Bruxelles, Bld du Triomphe, 1050 Brussels, Belgium}
\date{\today}

\preprint{ULB-TH/18-04}
\begin{abstract}
We investigate the phenomenology of scalar singlet dark matter candidates that
couple dominantly to the Standard Model via a Yukawa interaction with the top
quark and a colored vector-like fermion. We estimate the viability of this
vector-like portal scenario with respect to the most recent bounds from dark
matter direct and indirect detection, as well as to dark matter and vector-like
mediator searches at colliders. Moreover, we take QCD radiative corrections into
account in all our theoretical calculations. This work complements analyses
related both to models featuring a scalar singlet coupled through a vector-like
portal to light quarks, and to scenarios in which the dark matter is a Majorana
singlet coupled to the Standard Model through scalar colored particles (akin to
simplified models inspired by supersymmetry). Our study puts especially forward
the complementarity of different search strategies from different contexts,
and we show that current experiments allow for testing dark matter masses
ranging up to 700~GeV and mediator masses ranging up to 6~TeV.
\end{abstract}

\maketitle

\section{Introduction}

The hypothesis that dark matter (DM) may consist in weakly interacting massive
particles (WIMP) is currently being tested by various experiments including
direct and indirect DM probes, as well as colliders searches. In this
article,
we study a minimal setup in which a real scalar DM particle $S$ couples to the
Standard Model (SM) through interactions with a vector-like fermion. Such a
vector-like portal scenario has been the object of several previous studies,
which have focused on couplings either to light leptons~\cite{Toma:2013bka,%
Giacchino:2013bta,Giacchino:2014moa,Ibarra:2014qma} or  to light quarks~\cite{%
Giacchino:2015hvk}. A distinctive feature of this class of portals is that
radiative corrections tend to play a major role in DM annnihilation
phenomenology. In particular, virtual bremsstrahlung or annihilations into
mono-energetic photons and gluons may be the dominant mechanism driving the DM
relic abundance. By the same token, these may give rise to remarkable spectral
features, like a gamma-ray line that consists of a smoking gun for many DM
searches. For physical, but also technical reasons, previous studies have
nevertheless been limited to couplings to light SM leptons or quarks. In this
work we
complement these studies by considering a scenario in which the DM
particle solely couples, at tree-level, to the top quark through interactions
with a vector-like quark $T$. We explore different approaches to
investigate the phenomenological viability of the model from both collider and
cosmology standpoints.

In our predictions, we take into account several
higher-order corrections that include the QCD Sommerfeld effect and
next-to-leading-order (NLO) QCD corrections to the DM annihilation cross
section, both in the Early Universe and for what concerns indirect searches. For
the latter purpose, we have in particular computed the ${\cal O}(\alpha_s)$
corrections to the $S S \rightarrow t\bar t$ annihilation process, which
involves contributions from gluon emission by both the final-state top quarks as
well as by the virtual intermediate $t$-channel vector-like mediator. Although
the treatment of the associated infrared and collinear divergences is more
involved for heavy quarks than when the DM candidate is coupled to light
fermions, we only comment briefly on the associated difficulties and refer
instead to Ref.~\cite{Colucci:2018qml} and Ref.~\cite{Bringmann:2017sko} for details on
the scalar DM and Majorana DM cases respectively. We complement these constraints stemming from the relic density of DM
and its indirect detection null results by a study of the relevance of existing
direct DM probes. Our calculations take into account the effective coupling
of the dark scalar $S$ to gluons through loops involving top quarks and $T$
mediators~\cite{Hisano:2010ct}. 

On different lines, we estimate how collider
searches for both the vector-like partner $T$ and the DM particle $S$
restrict the model. We extend a previous study relying on simplified model
results from the Run~1 of the Large Hadron Collider (LHC)~\cite{Kraml:2016eti}
by considering more recent LHC Run~2 supersymmetry searches that can be recast
to constrain any model featuring strongly-interacting quark partners decaying
into a final-state comprised of missing energy and several SM objects~\cite{Chala:2017xgc}. We
moreover
include NLO QCD corrections through the computation of the corresponding matrix
elements and match the fixed-order predictions with parton showers~\cite{%
Fuks:2016ftf}, so that a state-of-the-art modeling of the LHC signals is used.
We additionally investigate the reach of the dedicated DM searches at
the LHC in the mono-X channels where the final-state signature consists in a
pair of DM particles recoiling against a very hard SM object X.

The plan of this article is as follows. In section \ref{sec:model} we define the
model and the associated parameter space. In section~\ref{sec:relic}, we discuss
our calculation of the DM relic abundance and how the latest results constrain
the parameter space. In section \ref{sec:astro_bounds} we further derive bounds
stemming from DM direct and indirect detection searches, and we finally address
the collider phenomenology of the model in section \ref{sec:lhc}. We emphasize
the complementarity of the different approaches in section \ref{sec:summary}, in
which we summarize the various cosmological and collider bounds that we have
obtained.

\section{Theoretical framework}
\label{sec:model}

We consider a simplified top-philic DM setup in which we extend the
Standard Model with a real scalar DM candidate $S$ with a mass $m_S$
and whose interactions with the Standard Model are mediated by exchanges of a
heavy vector-like quark $T$ of mass $m_T$. The $T$ quark is as usual considered
as lying in the fundamental representation of the QCD gauge group $SU(3)_c$, and
we focus on a minimal option where it is a weak isospin singlet with an
hypercharge quantum number set to 2/3. In order for the $S$ particle to be a
stable DM candidate, we impose a $\mathbb{Z}_2$ discrete symmetry under which
all Standard Model fields are even and the new physics states are odd.
Provided the $\mathbb{Z}_2$ symmetry is unbroken, it
prevents the $S$ field from mixing with the Standard Model Higgs doublet $\Phi$
and forbids the mixing of the $T$ quark with the Standard Model
up-type quark sector.

Our model is described by the Lagrangian
\begin{equation}\bsp
 {\cal L} =&\ {\cal L}_{\rm SM}
  + i \bar T \slashed{D} T - m_T \bar T T
 +  \frac 12 \partial_\mu S \partial^\mu S\\
 &\  - \frac 12 m_S S^2 + \Big[ \tilde{y}_t\, S\ \bar T P_R t + {\rm h.c.} \Big]
   -\frac12 \lambda S^2 \Phi^\dag \Phi\ ,
\esp\label{eq:lag}\end{equation}
where $P_R$ denotes the right-handed (RH) chirality projector and $t$ the top quark
field. The interaction strength between the mediator $T$, the DM and the SM
sector (or equivalently the top quark) is denoted by $\tilde{y}_t$. Like the DM, the vector-like mediator field $T$ is odd under $\mathbb{Z}_2$ but otherwise transforms  as the RH top field under $SU(3)\times SU(2) \times U(1)$ (and so has electric charge $Q = + 2/3$). 
A similar effective
Lagrangian has been considered in Ref.~\cite{Giacchino:2015hvk} in the
case of a DM particle coupling to light quarks, and, more recently, in
Refs.~\cite{Baek:2016lnv,Baek:2017ykw} for a coupling to the top
quark. In contrast to these last two studies, our analysis
differs in the treatment of the radiative corrections
that are relevant for the relic abundance, DM indirect and direct
detection as well as for the modeling of the collider signals.

The core of this work focuses on the phenomenological implications of the
presence of a colored vector-like $T$ particle mediating the interactions
of dark matter with the Standard Model. We therefore assume that the coupling of
the DM particle to the Higgs boson $\lambda$ appearing in the Lagrangian of
Eq.~\eqref{eq:lag} can be neglected, so that we set $\lambda = 0$. We moreover
impose that any loop-contribution to $\lambda$ could be absorbed in the
renormalization procedure and thus ignored. Details on
departures from this hypothesis can be found in Ref.~\cite{Baek:2016lnv}.
This contrasts with the analogous case in which the dark matter particle
consists in a Majorana fermion that couples to the SM top quark through a scalar
colored mediator, as in the latter new physics setup, an effective DM-Higgs
coupling arises at the one-loop level, is calculable and
finite~\cite{Garny:2018icg}.

The relevant model parameter space is therefore defined by three parameters,
namely the two new physics masses $m_T$ and $m_S$, and the Yukawa
coupling $\tilde{y}_t$.

\section{Dark matter relic density}
\label{sec:relic}

\subsection{Radiative corrections}
\label{sec:bremsstr-radi-corr}

It has been recently shown that radiative corrections to the DM annihilation
cross section play a significant role in the phenomenology of a real scalar DM
candidate, either through internal bremsstrahlung or via new channels that open
up (like for instance when DM annihilates into a pair of
monochromatic gluons or photons)~\cite{Giacchino:2013bta,Toma:2013bka,%
Giacchino:2015hvk,Giacchino:2014moa,Ibarra:2014qma}. All these analyses have
however been restricted to scenarios featuring a DM particle coupling
to light SM
quarks or leptons, so that the corresponding fermion masses could be neglected
and the calculation performed in the so-called chiral limit. When non-vanishing
SM fermion masses are accounted for, the computation of the radiative
corrections to the annihilation cross section is plagued by infrared
divergences that must be consistently handled, as it has been studied in details
for Majorana DM~\cite{Bringmann:2015cpa}. The scalar DM case has been thoroughly
analyzed by some of us~\cite{Colucci:2018qml}, so that we summarize in this section the
points that are the most relevant for our study.

The calculation of the annihilation cross section associated with the
$S S \rightarrow t\bar t$ process at ${\cal O}(\alpha_s)$ involves
contributions both from final state radiation (FSR) and from virtual internal
bremsstrahlung (VIB) diagrams. The corresponding amplitudes exhibit a specific
dependence on the kinematics, which reflects in distinguishable features in the
spectrum of radiated photons or gluons. In particular, VIB tends to yield a
final-state energy spectrum that peaks at high energies $E_{\gamma,g} \lesssim
m_S$. Whilst FSR contributions also lead to the emission of a hard gluon or
photon~\cite{Birkedal:2005ep}, the related spectral feature is less remarkable
than in the VIB case, unless VIB is relatively suppressed. For a fixed
DM mass $m_S$, the relative FSR and VIB weights are controlled by the mass of
the vector-like mediator $m_T$ and by the final state quark mass ({\it i.e.},
the mass of the top quark $m_t$). FSR turns out to be less important as
$m_t/m_S$ decreases, since the contribution to the annihilation cross section is
proportional to the leading order (LO) result which, in an $s$-wave
configuration, is helicity suppressed. In the chiral limit, $m_t/m_S \to 0$ and
FSR can thus be neglected. On the other hand, the VIB spectral features are
controlled by the $m_T/m_S$ mass ratio, and the energy spectrum peaks toward
$E_{\gamma,g} \sim m_S$ as $T$ and $S$ become more mass-degenerate.

For generic particle masses, both FSR and VIB features are present and must be
accounted for. This consequently requires a consistent handling of the infrared
and collinear divergences of the FSR amplitude, that are only cancelled out
after including the virtual contributions as guaranteed by the Kinoshita-Lee-%
Nauenberg theorem. The associated computations are facilited when
carried out in an effective approach (with a contact $SS t\bar t$ interaction)
that suits well for the annihilation of non-relativistic DM particles in the
soft and collinear limit~\cite{Colucci:2018qml,Bringmann:2015cpa}. The hard part of the
spectrum is then described by the $SS \rightarrow t \bar t g$ (or $t\bar t
\gamma$) contribution as calculated from the full theory of Eq.~\eqref{eq:lag},
and the two results are matched by using a cutoff on the energy of the radiated
gluon (photon). This approach allows us to get a
regularized expression for the total $SS$ annihilation cross section at the NLO
accuracy that is valid for a broad range of parameters~\cite{Colucci:2018qml}.

The procedure outlined above vindicates the fact that for a large part of the
parameter space, one may rely on a simple approximation for the total
annihilation cross section,
\begin{equation} \label{eq:svttgall}
  \renewcommand{\arraystretch}{1.3}
  \sigma v_{t\bar t} |_{\rm NLO} \approx
  \left\{
   \begin{array}{ll}
       \sigma v_{t\bar t} & m_S<300 \,{\rm GeV,}\\
    \sigma v_{t\bar tg}\vert_{m_t=0}+\sigma v_{t\bar t} &m_S>300 \,{\rm GeV.}
   \end{array} \right.
\end{equation}
In this expression, $\sigma v_{t\bar t}$ is the $s$-wave contribution to the LO
annihlation cross section,
\begin{equation}
 \label{eq:SStoqq}
\sigma v_{t \bar t}^{s{\rm-wave}}  = \frac{3 \tilde{y}_t^4 }{4 \pi m_S^3} \frac{m_t^2\, (m_S^2 - m_t^2)^{3/2}}{(m_S^2+m_T^2- m_t^2)^2}  \ ,
\end{equation}
and $\sigma v_{t\bar tg}\vert_{m_t=0}$ is the ($s$-wave) annihilation
cross section as obtained in the chiral limit and when a single gluon
radiation is included. Its explicit form can be found in
Refs.~\cite{Ibarra:2014qma,Giacchino:2013bta,Colucci:2018qml}. The difference with the
exact result is only large for $m_S \simeq m_t$ and reaches at most
30\% beyond (see Fig.~\ref{fig:svtt-g-ratios} discussed in the
framework of section~\ref{sec:parameter-space}). When $m_S \rightarrow
m_t$, the treatment used for the derivation of $\sigma v_{t\bar tg}$
breaks down due to threshold corrections that affects the production
of a top-antitop system nearly at rest~\cite{Drees:1990dq}, an artefact that is
visible in Fig.~\ref{fig:svtt-g-ratios} in the region below $m_S$ of about
300~GeV. For $m_S \sim 300$ GeV,
$\sigma v_{t\bar t}|_{\rm NLO}\approx\sigma v_{t\bar t}$.
Whereas the procedure allowing to deal with threshold effects relevant for this
mass configuration is in principle well-know~\cite{Drees:1989du}, its
implementation goes beyond the scope of our work. Those effects not
only concern a narrow range of parameters, but in the absence of toponium bound
states, they are also expected to yield small and sub-leading corrections to the
LO annihalition cross section~\cite{Colucci:2018qml}. For this reason, the LO
annihilation cross section $\sigma v_{t \bar t}$ is used for scalar masses below
about 300~GeV. For larger scalar masses, we additionally include the
contribution of internal bremsstrahlung, calculated in the massless quark limit.
Such an approximation provides a smooth transition to the mass regime in which
gluon emission consists in the dominant contribution to the annihilation cross
section, {\it i.e.}, for $m_S$ of a few TeV~\cite{Colucci:2018qml}.

\subsection{Relic abundance}
\label{sec:parameter-space}
\begin{figure*}
  \begin{center}
    \hspace*{-1.cm}
    \begin{tabular}{cc}
     \includegraphics[width=8cm]{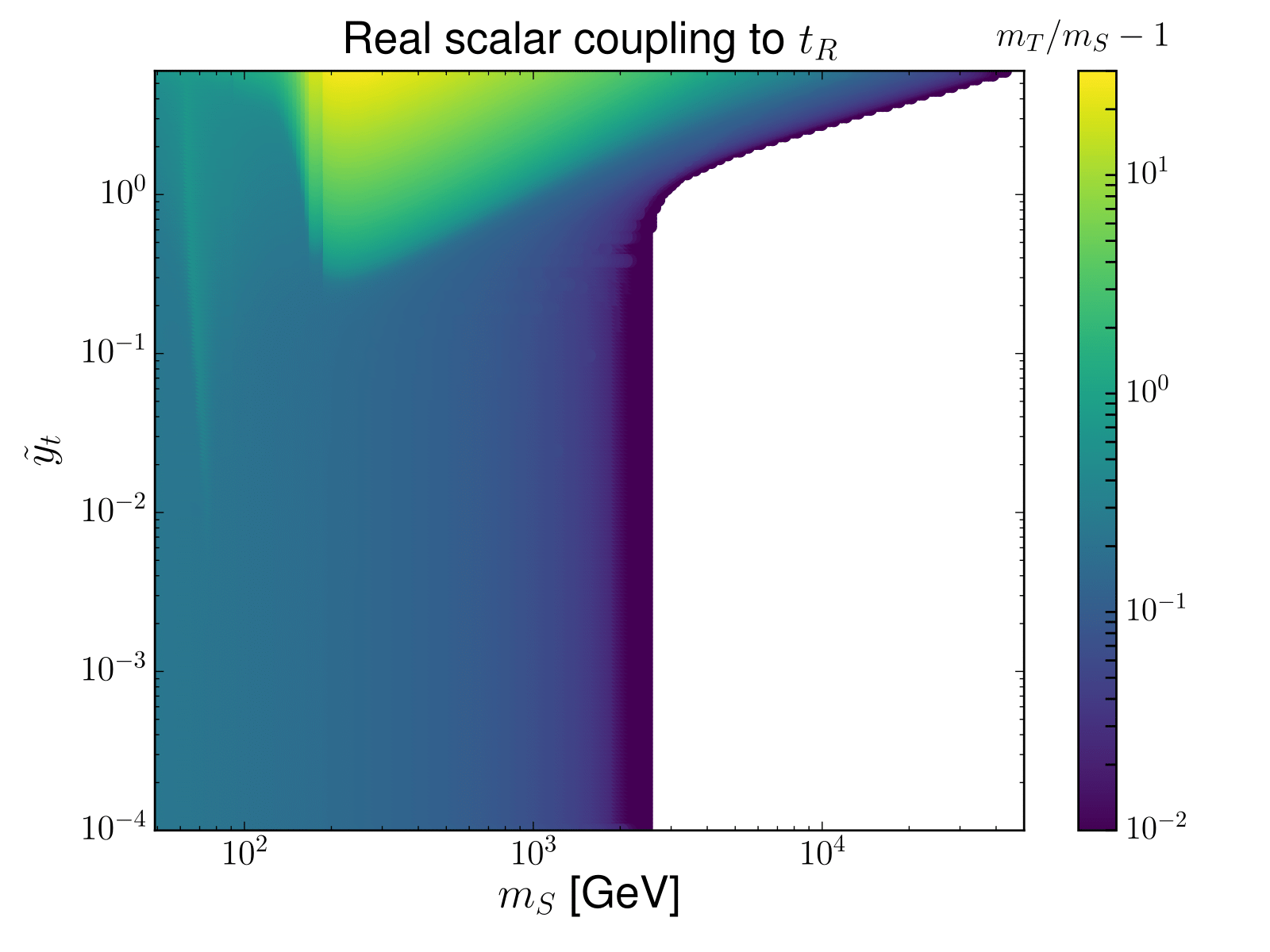}&
     \includegraphics[width=8cm]{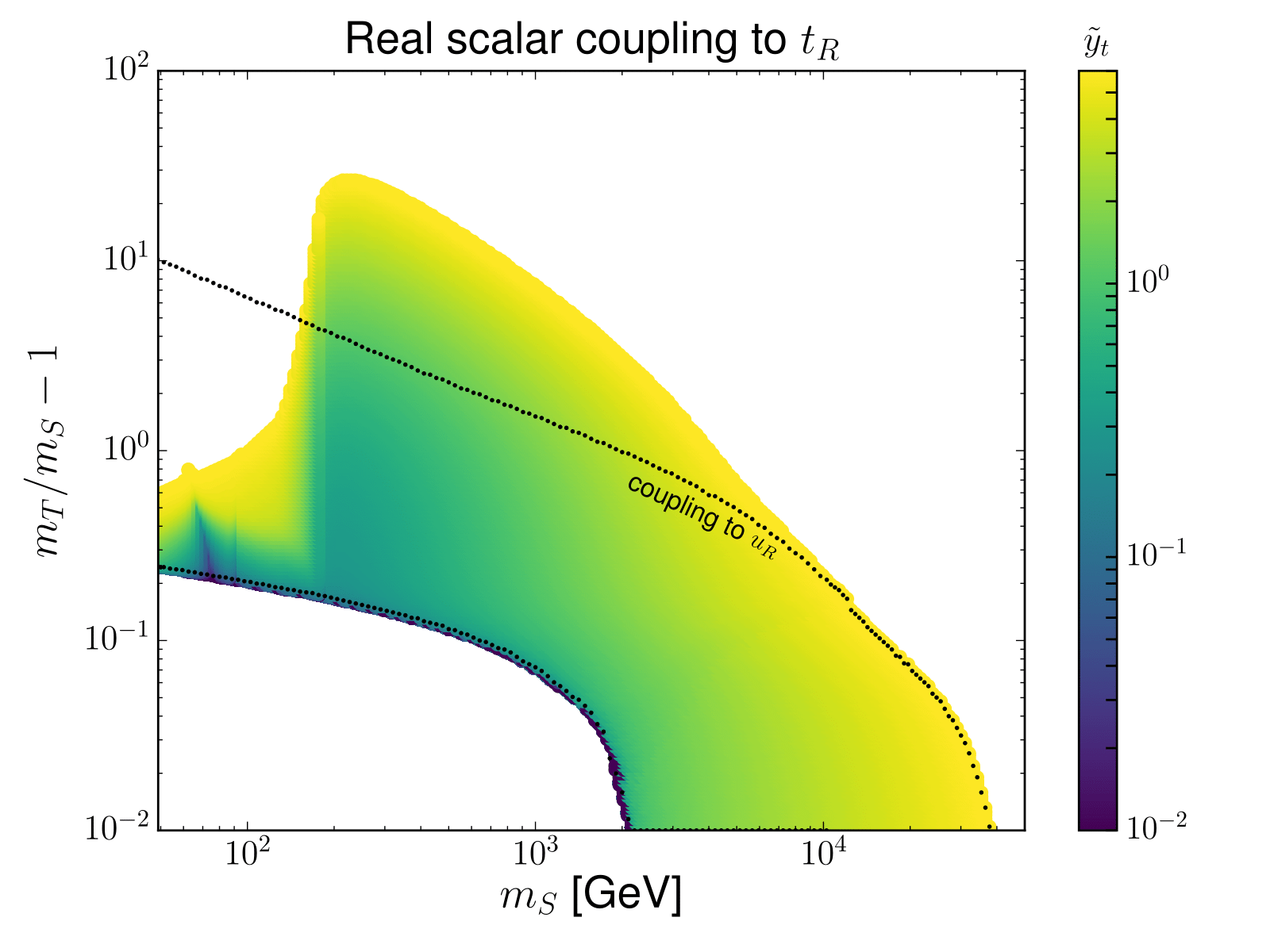}
     \end{tabular}
      \end{center}
  \caption{Region of our parameter space for which one can accommodate a relic
    abundance of $\Omega h^2=0.12$. The results are shown in the $(m_S,
    \tilde{y}_t)$ plane (left) and $(m_S,r-1)$ plane (right), the color code
    being associated with the value of the $r-1$ and $\tilde{y}_t$ parameters
    respectively. For comparison, the dotted black contour in the right panel
    represents the expected parameter space coverage in the case of a scalar DM
    particle coupling to right-handed up quarks $u_R$.}
\label{fig:viab}
\end{figure*}

In order to determine the relic abundance of the dark $S$ particle, we
consider the freeze-out mechanism for DM production in the Early
Universe and make use of the {\sc MicrOMEGAs}
code~\cite{Belanger:2014vza}, which we have modified in order to
accommodate some of the particularities of our model. These include dark
matter annihilations into a $tWb$ three-body final state once one lies
below the top threshold, the radiative corrections mentioned in
section~\ref{sec:bremsstr-radi-corr} and Sommerfeld effects. The
latter especially affect vector-like fermion annihilation and dark
matter co-annihilation with a mediator, those corrections contributing
to the relic abundance by at most 15\% (see
appendix~\ref{sec:somm-corr}). In addition, DM annihilations
into a $tWb$ system play a non negligible role for DM masses lying in
the $[(m_t+m_W)/2, m_t]$ mass window, and we have included these
contributions by evaluating them numerically with {\sc
  CalcHEP}~\cite{Belyaev:2012qa}. Finally, we have added the
  loop-induced $SS\to gg$ and $SS\to\gamma\gamma$ processes in the
  computation of the DM annihilation cross section
  ~\cite{Giacchino:2014moa,Ibarra:2014qma}. The annihilation into
  gluons is in particular significant for DM masses below the
  top threshold.

We present the results in Fig.~\ref{fig:viab}, under the form of two
two-dimensional projections of our three-dimensional parameter
space. In the left panel of the figure, we show the region of the
$(m_S, \tilde{y}_t)$ plane for which there exists a mediator mass
value yielding to a relic density $\Omega h^2=0.12$ compatible with the Planck
results~\cite{Ade:2013zuv}.
The gradient of colors in Fig.~\ref{fig:viab} is associated to
relative mass difference between the DM and the mediator given by
$r-1$ with
\begin{equation}
   r = \frac{m_T}{m_S} \, .
\end{equation}
Similarly, we present in the right panel of the Fig.~\ref{fig:viab} the region
of the $(m_S,r-1)$ plane for which there exists a $\tilde{y}_t$ coupling value,
shown through a color code, leading to the observed relic abundance. The Yukawa
coupling is enforced to lie in the $[10^{-4}, 6]$ window, the upper bound being
an extreme value at the limit of the
perturbative regime (defined by $\tilde y_t g_s/4\pi < 1$) and the lower bound
guaranteeing the correct treatment of the co-annihilation processes by {\sc
MicrOMEGAs}. For $\tilde{y}_t > 10^{-4}$, co-annihilation processes like $St \to
Tg$ occur in chemical equilibrium, and the DM abundance is determined by a
single Boltzmann equation involving an effective annihilation cross section
accounting for co-annihilations~\cite{Edsjo:1997bg}. For smaller $\tilde{y}_t$
values, thermal freeze-out could still yield the observed DM abundance, but a
larger system of Boltzmann equations involving the abundance of both the $T$ and
$S$ particles has to be accounted for in order to precisely determine the
departure from chemical equilibrium~\cite{Garny:2017rxs,Garny:2018icg}. This
issue is left for a possible future work.

The two panels of Fig.~\ref{fig:viab} provide complementary information.  In the
$(m_S, \tilde y_t)$ plane, one observes parameter space regions in which the DM
abundance is driven by co-annihilation processes and so feature little
dependence on the $\tilde y_t$ value. They correspond to setups for which
$m_T/m_S-1$ is of at most ${\cal O}(0.1)$, and which are represented by the thin
dark blue region in the complementary $(m_S, m_T/m_S-1)$ plane. For the sake of
the comparison, we also superimpose in
Fig.~\ref{fig:viab} the limits of the viable parameter space (black
dotted contour) of a model for which the DM couples to the
right-handed up quark $u_R$. We refer to Ref.~\cite{Giacchino:2015hvk}
for more details.

The viable part of the parameter space can be divided into three distinct
regions according to the DM mass $m_S$.

\begin{figure}
      \includegraphics[width=8cm]{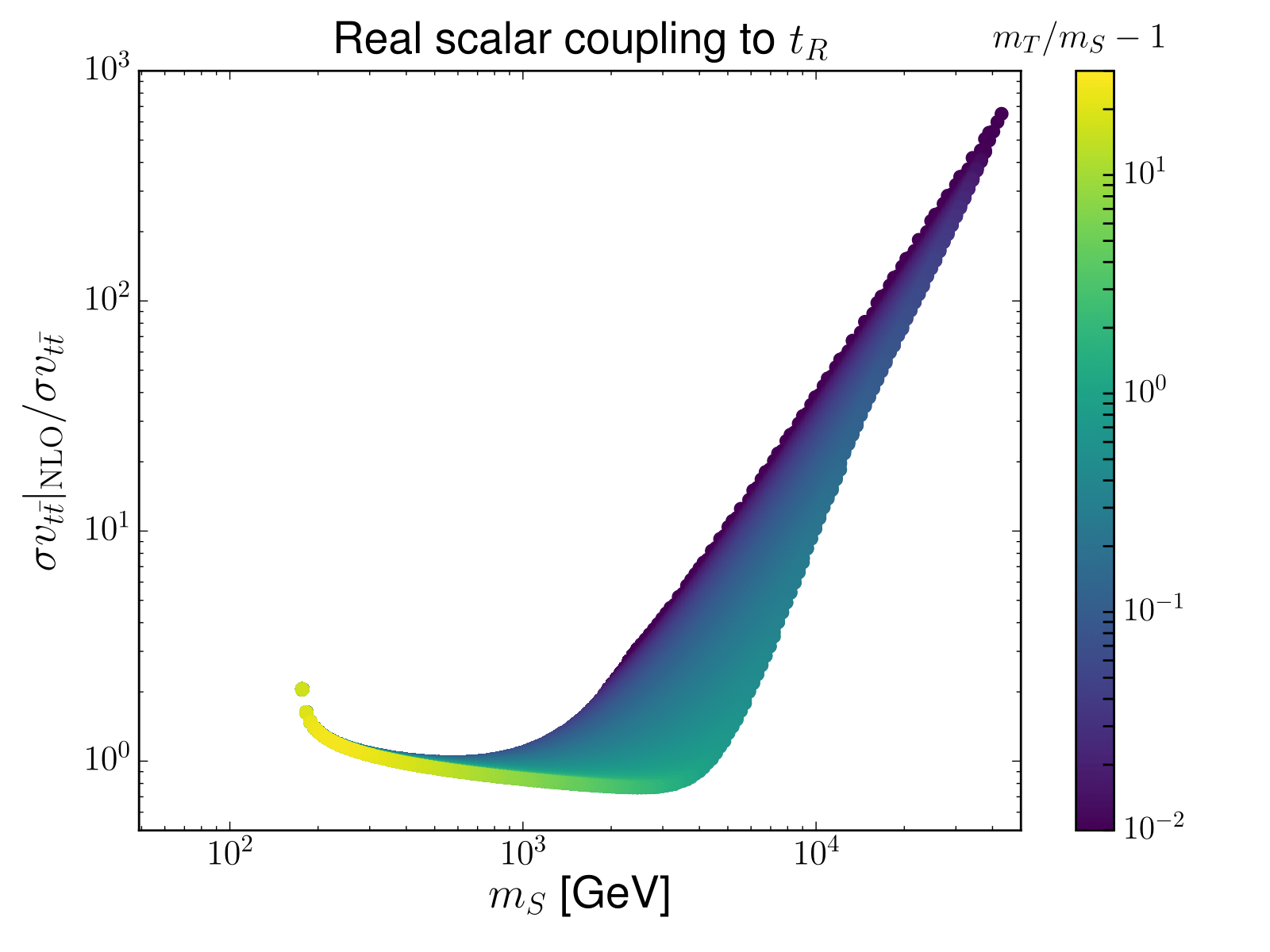}
 \caption{Ratio of the exact NLO DM annihilation cross section
   $\sigma v_{ t \bar tg}|_{\rm NLO}$~\cite{Colucci:2018qml} to the two-body LO cross
   section $\sigma v_{t\bar t}$. This shows that gluon radiation consists in the
   dominant component of the annihilation cross section for DM masses
   satisfying $m_S \gtrsim 5$~TeV. 
   In the figure, all points correspond to models matching the correct
   DM abundance and the color code represents the value of $r-1$.}
\label{fig:svtt-g-ratios}
 \end{figure}

\noindent $\bullet$ \underline{$\mathbf{m_S>5}$ {\bf TeV}}. For very
heavy DM, the mass of the top quark only plays a subleading role. This
is clearly visible in the right panel of Fig.~\ref{fig:viab}, where
the viable region of the parameter space of the top-philic scenario
matches the one expected in the up-philic case. In this regime,
$m_S\gg m_t$ and the chiral limit approximation for the DM
annihilation cross section is valid.  Moreover, VIB
corrections are large, as illustrated in Fig.~\ref{fig:svtt-g-ratios}
where we show, for all benchmark points giving rise to the right DM
abundance in Fig.~\ref{fig:viab}, the ratio of the exact NLO
result~\cite{Colucci:2018qml} to the LO predictions $\sigma v_{t\bar t}$.
 The importance of the
NLO corrections will be further discussed in the context of DM
indirect detection bounds in section~\ref{sec:indirect}.

\vspace{.5cm}
\noindent $\bullet$ \underline{$\mathbf{m_t <m_S <5}$ {\bf TeV}}. In
this regime where the DM mass is moderate, the tree-level $s$-wave
$SS\to t\bar t$ contribution to the annihilation cross section
dominates, as additionally illustrated in Fig.~\ref{fig:svtt-g-ratios}
where the NLO to LO ratio is close to 1. Notice that the feature
observed for $m_S \sim m_t$ in Fig.~\ref{fig:svtt-g-ratios} is
spurious as correct predictions must include threshold effects that we
have ignored. The LO annihilation into a pair of quarks is, in
contrast, completely negligible in the light quark case for which the
relic density is driven by loop-induced annihilations into
gluons~\cite{Giacchino:2015hvk}. The phenomenologically viable region of
the parameter space in the top-philic scenario consequently strongly
deviates from the corresponding one in the up-philic model, as shown
in the right panel of Fig.~\ref{fig:viab}.  Given that finite quark
mass effects are significant, larger $r$ parameters are found
acceptable for a given DM mass in the top-philic case.

\noindent $\bullet$ \underline{$\mathbf{m_S < m_t}$}. In this regime,
the DM abundance is driven either by annihilations into a $tWb$ system
via a virtual top quark (for $m_S \lesssim m_t$),  through
loop-induced annihilations into gluons, or through co-annihilations
with the mediator. Any other potential contribution, like DM
annihilations into pairs of SM particles through the Higgs portal (as
it occurs in the scalar singlet DM
scenario~\cite{Cline:2013gha,Athron:2017kgt}) is here irrelevant since we have set
the $\lambda$ quartic coupling in Eq.~\eqref{eq:lag} to
zero. Co-annihilations particularly play an important role near
$m_T+m_S\simeq m_t$, as the $ST\to t\to tg$ channel is resonantly
enhanced.  This corresponds to the light-yellow region in the left
panel of Fig.~\ref{fig:viab} for $m_S\sim70-80$ GeV, and to the blue
peak in the right panel of the figure for the same $m_S$
values. Annihilations into monochromatic gluons are only important
when the mass of the mediator is large enough to close all
co-annihilation channels, and annihilations into a $tWb$ three-body
system are only relevant close to threshold, for $m_S\in [(m_t+m_W)/2,
  m_t]$.

\section{Direct and indirect constraints}
\label{sec:astro_bounds}

 \subsection{Direct detection constraints}
\label{sec:direct-detect-constr}
\begin{figure}
  \begin{center}
     \includegraphics[width=.32\columnwidth]{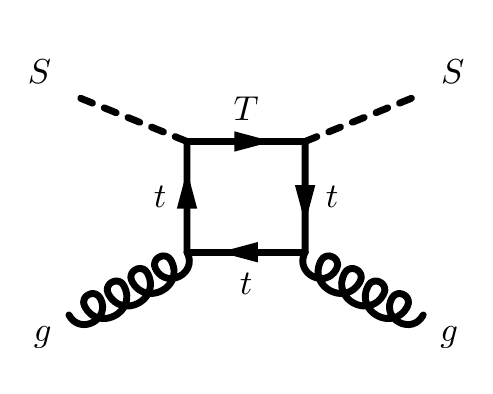}
     \includegraphics[width=.32\columnwidth]{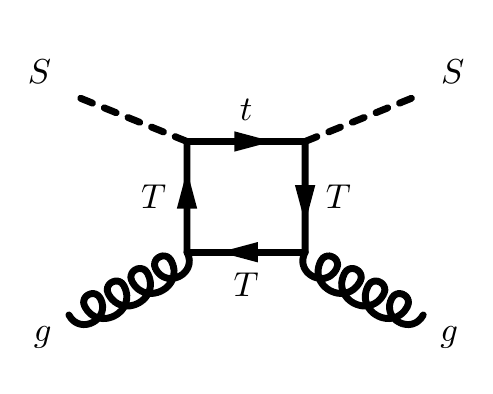}
     \includegraphics[width=.32\columnwidth]{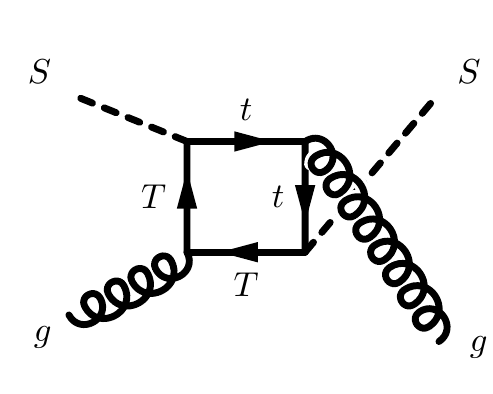}
  \end{center}
  \caption{Feynman diagrams relevant for DM-nucleon scattering.}
\label{fig:diagDD}
\end{figure}

In the limit in which the quartic coupling of $S$ to the Higgs boson vanishes,
the DM nucleon scattering cross-section can be computed from the evaluation of
the one-loop diagrams shown in Fig.~\ref{fig:diagDD}. This allows one to derive
an effective Lagrangian for the DM coupling to gluons,
\be
 \mathcal{L}_{g} = C_S^g \, \frac{\alpha_s}{\pi} S^2 \, G^{\mu \nu} G_{\mu \nu} \,, \label{eq:Lg_eff} 
\end{equation}
where the Wilson coefficient $C_S^g$ includes both short and long-distance
contributions (relatively to the momentum scale involved in the
loop)~\cite{Hisano:2010ct,Gondolo:2013wwa}. The resulting effective
spin-independent coupling $f_N$ of the scalar DM particle $S$ to a
nucleon $N$ of mass $m_N$ is then given by~\cite{Drees:1993bu}
\begin{equation}
  \frac{f_N}{m_N} =  - \frac89 C_S^g f_{T_G}^{(N)}
  \ \  \text{with}\ \
  f_{T_G}^{(N)}=1-\sum_{q=u,d,s}f_{T_q}^{(N)} \ ,
\label{eq:fN} \end{equation}
where the quark mass fractions $f_{T_q}^{(N)}$ and the analytical expression for
$C_S^g$ can be found in Ref.~\cite{Hisano:2015bma}.

We compute the total spin-independent cross section $\sigma_A$ for DM
scattering off a nucleus with charge $Z$ and a mass number $A$ by taking the
coherent sum of the proton and the neutron contributions,
\begin{equation}
  \sigma_A = \frac{m_A^2}{\pi (m_S+m_A)^2} \bigg[ Z f_p + (A-Z)f_n \bigg]^2 \,,
\label{eq:sigma_general}
\end{equation}
where $f_p$ and $f_n$ denote the respective DM couplings to a proton and a
neutron derived from Eq.~\eqref{eq:fN} with $N=p$ or $n$, respectively, and
$m_A$ is the nucleus mass.

\begin{figure}
  \centering
  \includegraphics[width=.98\columnwidth]{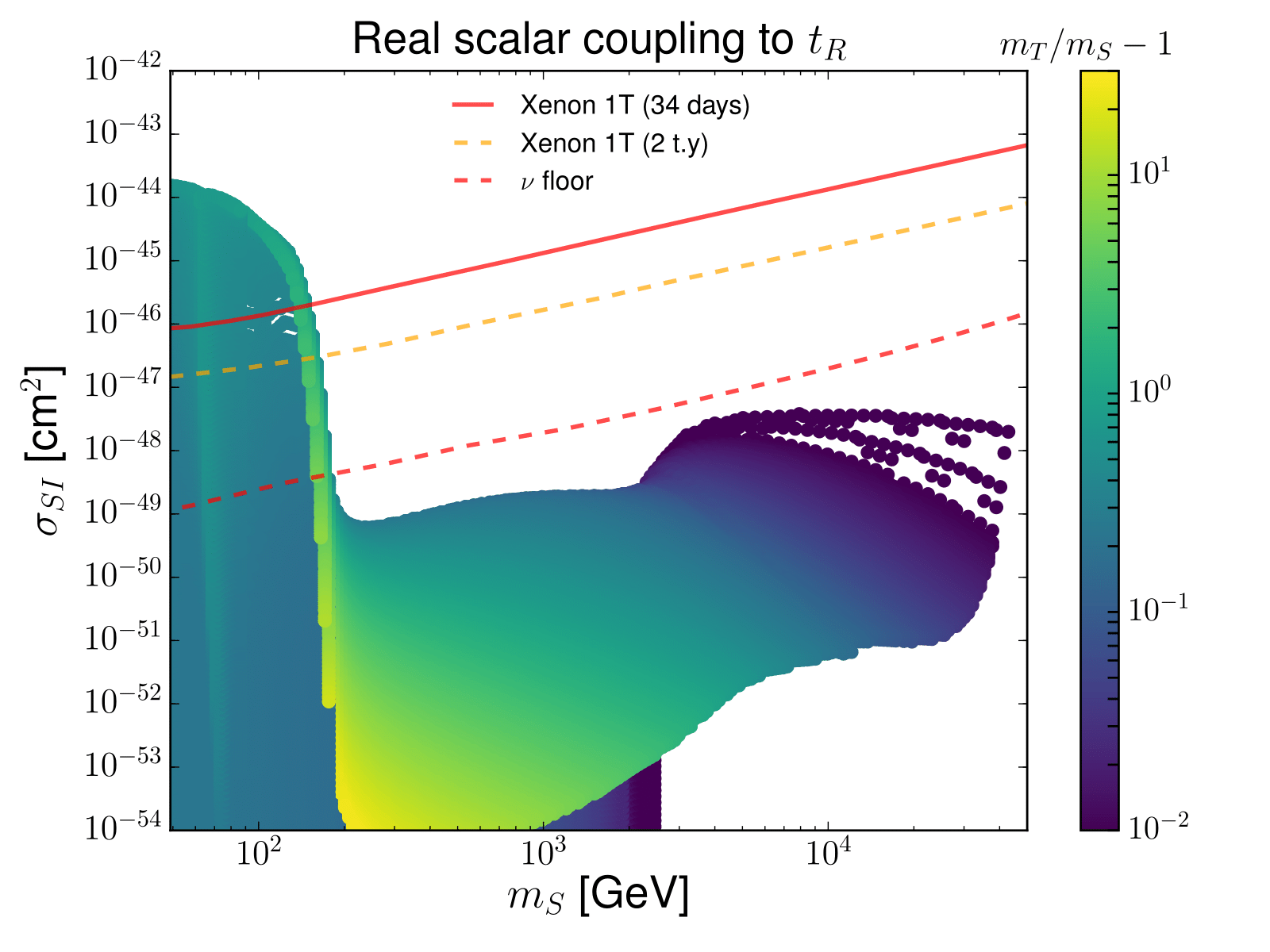}
  \caption{DM-proton spin-independent scattering cross section as a function of
    the DM mass $m_S$. For each scenario, the coupling to the top quark and the
    value of the $r-1$ parameter (shown trough the color code) are fixed
    to reproduce the observed relic density. The continuous red line represents
    the 90\% confidence level exclusion of the Xenon 1T experiment~\cite{%
    Aprile:2017iyp}, the orange dashed line the Xenon 1T reach~\cite{%
    Aprile:2015uzo} and the red dashed line the neutrino
    floor~\cite{Billard:2013qya}.}
\label{fig:DD1}
\end{figure}

In Fig.~\ref{fig:DD1}, we present the dependence of the DM scattering
cross section on protons $\sigma_{SI}$ calculated as depicted above, for all DM
scenarios of Fig~\ref{fig:viab}. For $m_S\lesssim m_t$, the models
featuring the largest $\sigma_{SI}$ values are those with the largest
$\tilde{y}_t$ value and for which the relic density is driven by
annihilations into a pair of gluons. As in the left panel of
Fig.~\ref{fig:viab}, the yellow region around $m_S \sim 80$~GeV
corresponds to scenarios for which resonant co-annihilations of the
$S$ and $T$ particles into a top quark play a leading role. Above the
top mass threshold, the Yukawa coupling required to match a correct
relic abundance drops, and so does the elastic scattering cross
section. The figure finally exhibits a bump above $m_S \gtrsim
2.5$~TeV, which corresponds to setups in which $m_S + m_t \sim
m_T$. The $C_S^g$ coefficient is then consequently enhanced, which
directly impacts the elastic cross section~\cite{%
  Hisano:2015bma}.

For most DM models $\sigma_{SI}$ lies however below the neutrino
floor, except for some scenarios with a DM candidate lighter than the
top quark. The constraints originating from the results of the
Xenon 1T experiment after 34 days of exposure~\cite{Aprile:2017iyp}
are also indicated, together with predictions under the assumption of
2.1 years of data acquisition~\cite{Aprile:2015uzo}. Although a large
part of the parameter space region lying above the neutrino floor is within the range of Xenon~1T, a
significant fraction of it will stay unconstrained in the near future
by DM direct detection searches. The corresponding excluded region
projected in the $(m_S, r-1)$ plane is presented in the summary of
Fig.~\ref{fig:summary}, after
accounting for the latest bounds from the Xenon 1T
experiment (red region), together with the region
that could be tested up to the neutrino floor (red dashed contour).

\subsection{Indirect detection constraints}
\label{sec:indirect}

\begin{figure}
  \begin{center}
    \includegraphics[width=0.91\columnwidth]{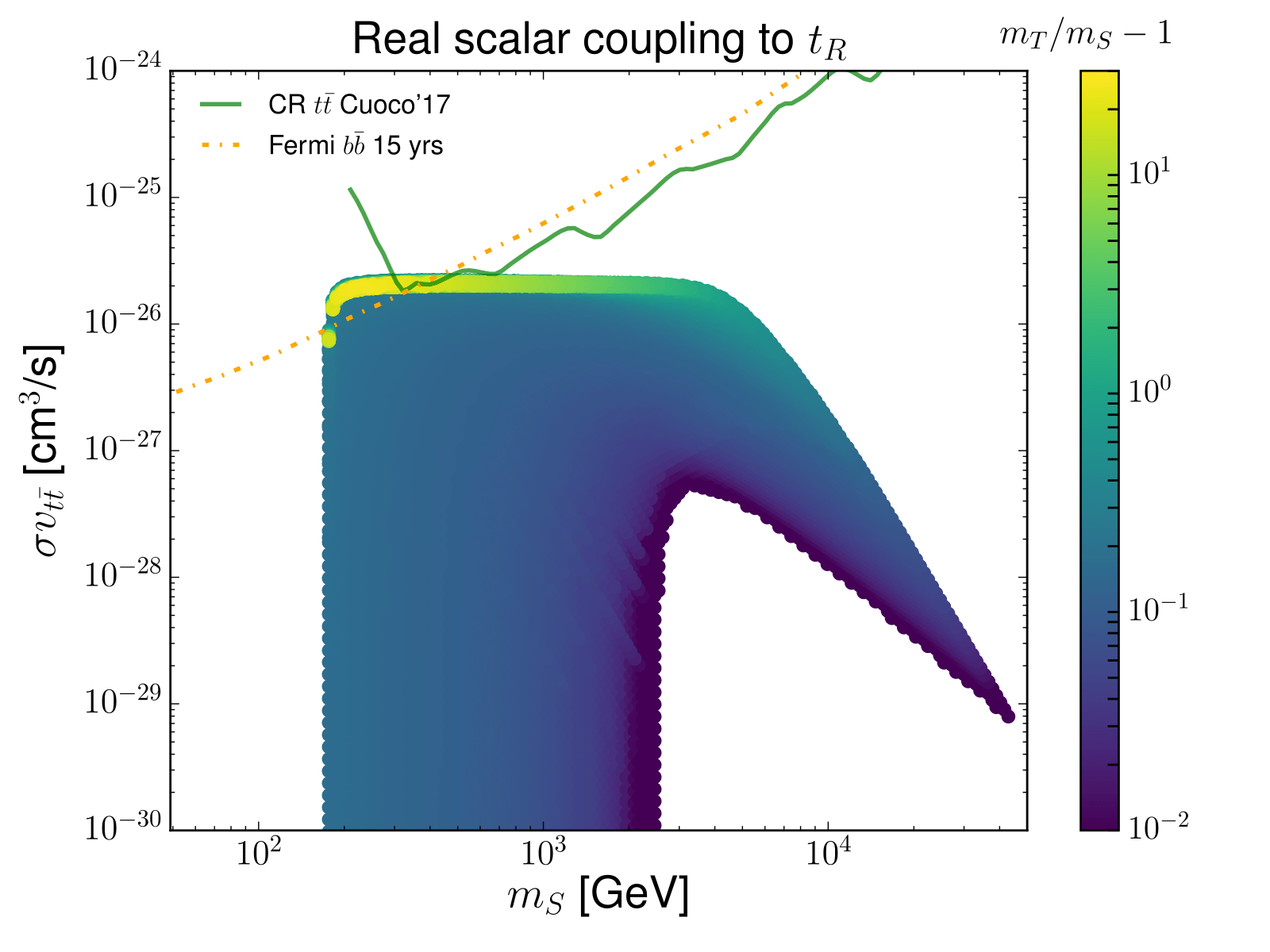}\\[-.1cm]
    \includegraphics[width=0.91\columnwidth]{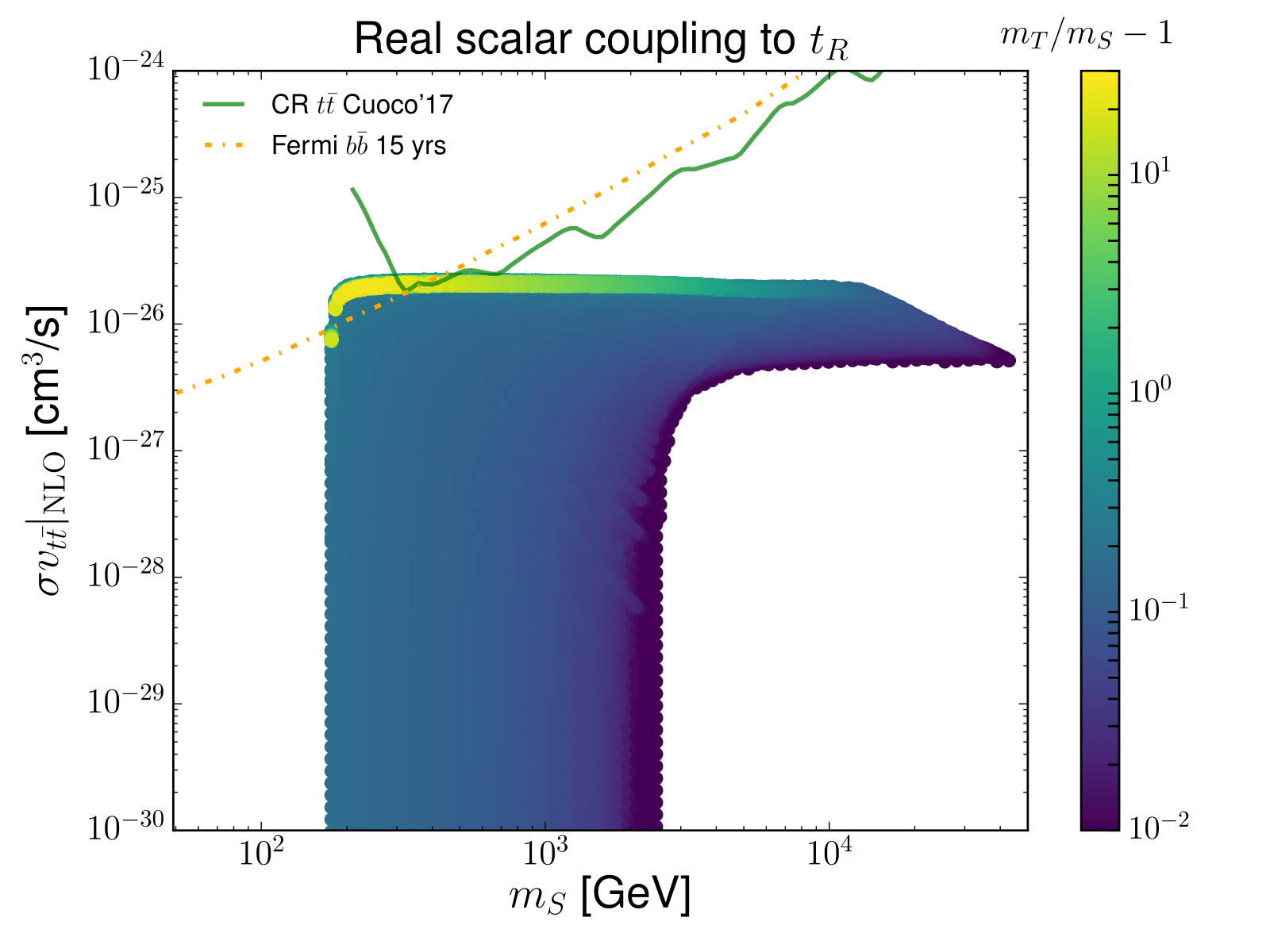}\\[-.1cm]
    \includegraphics[width=0.91\columnwidth]{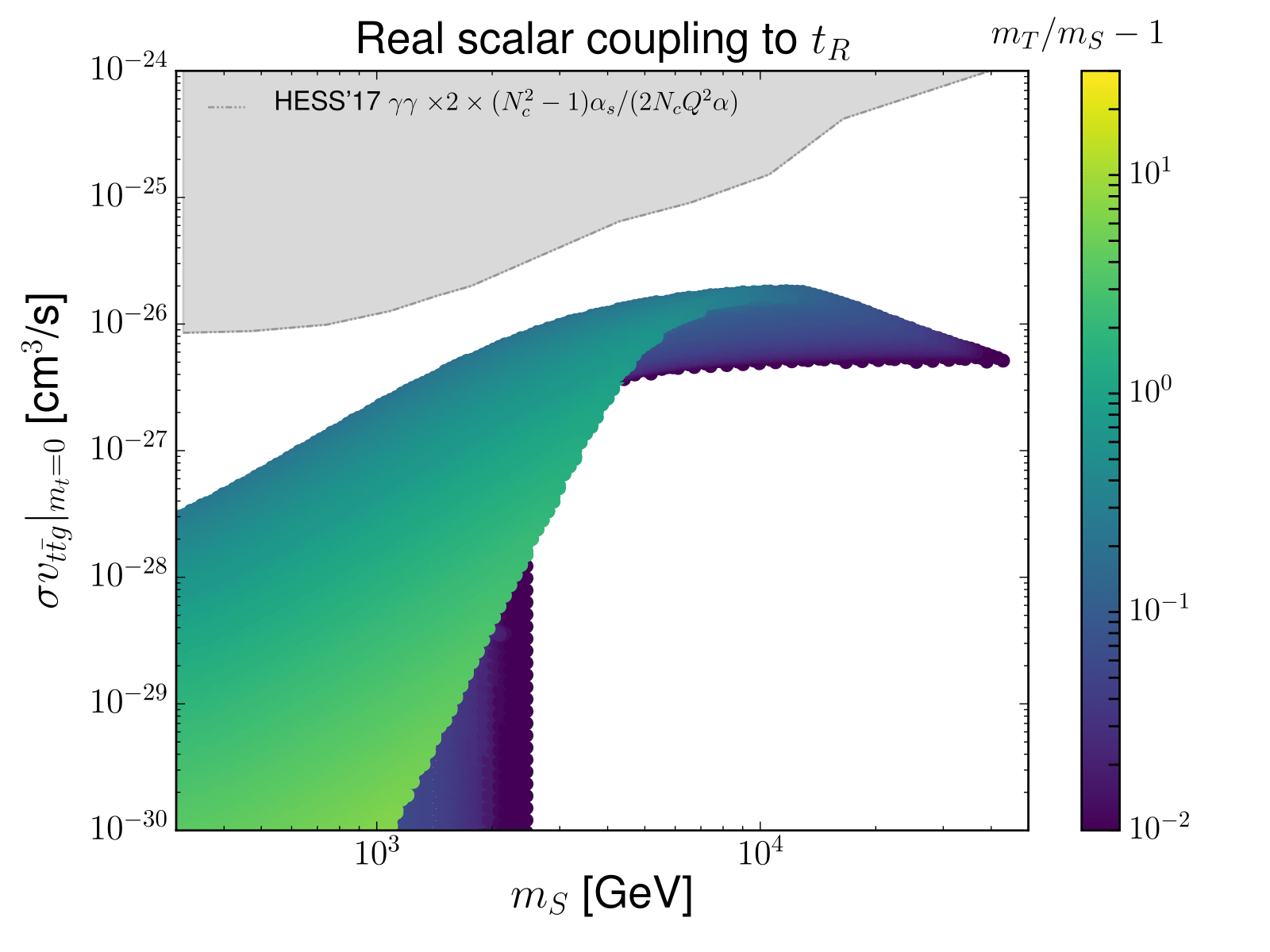}
    \caption{LO (upper panel) and NLO (central panel) $SS \to t\bar t$
      annihilation cross sections at zero velocity, where the NLO
      results are evaluated using the approximation of
      Eq.~\eqref{eq:svttgall}, as well as the $SS\rightarrow \bar t t
      g$ annihilation cross section in the chiral
      limit~\cite{Giacchino:2013bta} (lower panel). We superimpose to
      our results the indirect detection limits obtained from the
      cosmic ray (CR) analysis of Ref.~\cite{Cuoco:2017iax} (green
      continuous line), as well as the bounds that could be expected
      after 15 years of Fermi-LAT running when dwarf spheroidal galaxy
      data in the $b\bar b$ channel is analyzed~\cite{%
        Charles:2016pgz} (dot-dashed orange line). An
      estimation of the upper limits expected from gamma-ray line
      H.E.S.S. data~\cite{Rinchiuso:2017kfn} is also presented (see
      the text for details) in the lower panel (gray). The color code
      represents the value of the $r-1$ parameter and we have
      considered DM models satisfying the relic density
      constraints of section~\ref{sec:relic}.
    \label{fig:svtt-g}}
  \end{center}
\end{figure}

\begin{figure}
  \begin{center}
    \includegraphics[width=0.98\columnwidth]{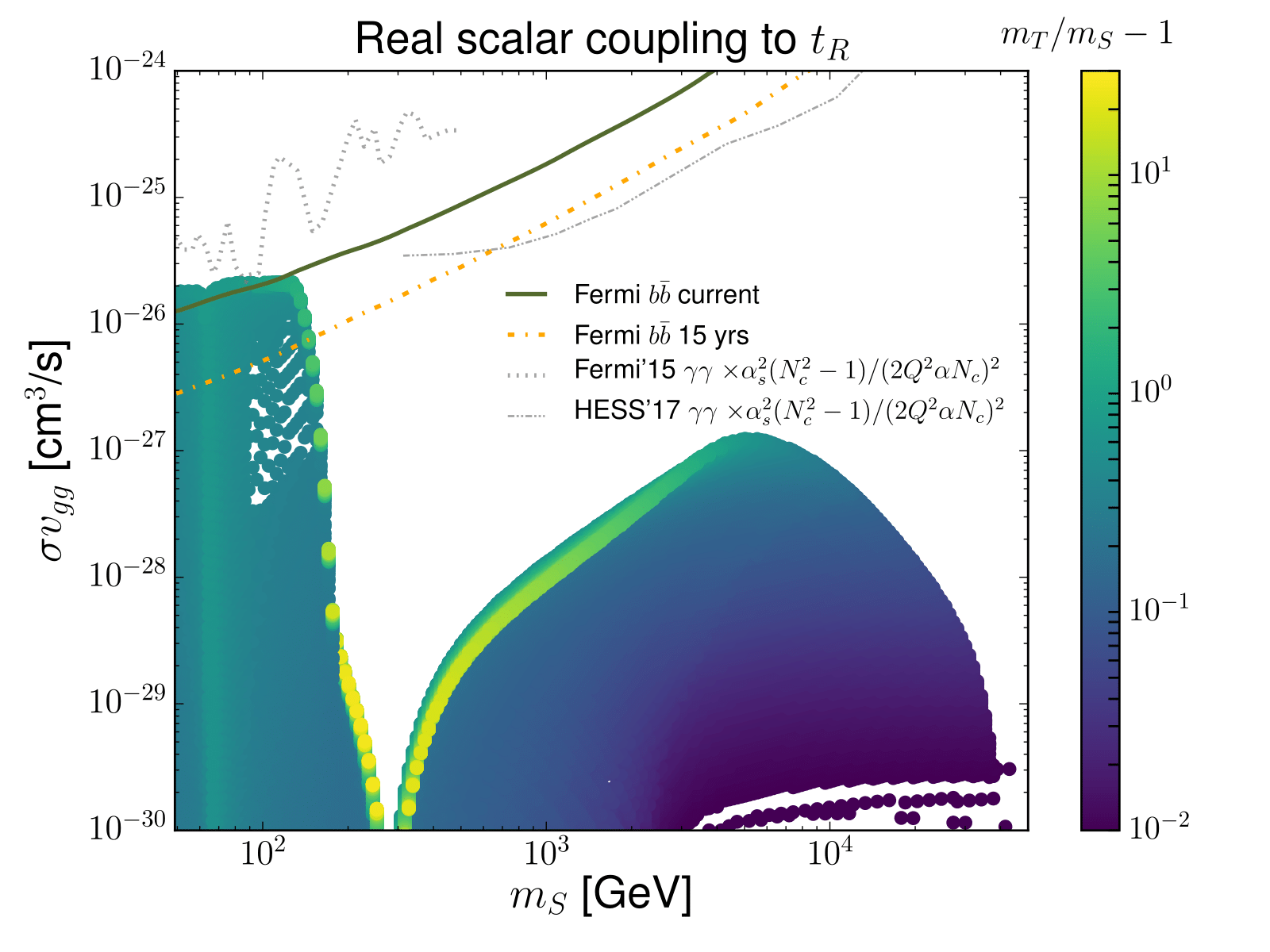}\\
  \caption{Predictions for the $SS\rightarrow gg$ annihilation cross
    section, to which we superimpose the upper limits extracted from
    the cosmic ray analysis of dwarf spheroidal galaxy data in the
    $b\bar b$ channel from Fermi-LAT~\cite{Ackermann:2015zua} using
    current results (dark green continuous line) and projected results
    assuming 15 years of data acquisition (orange dot-dashed
    line)~\cite{Charles:2016pgz}.  We also indicate the upper limits
    obtained from the gamma-ray line analysis of
    Fermi-LAT~\cite{Ackermann:2015lka} and
    H.E.S.S.~\cite{Rinchiuso:2017kfn} by gray dotted and double-dot-dashed
    lines, respectively (see the text for details). The color code
    represents the value of the $r-1$ parameter and we have
    considered DM models satisfying the relic density constraints
    of section~\ref{sec:relic}.
    \label{fig:svg}}
  \end{center}
\end{figure}

In Figs.~\ref{fig:svtt-g} and \ref{fig:svg}, we present, for all scenarios
satisfying the relic density constraints of section~\ref{sec:relic}, the value
of the DM annihilation cross section at zero velocity into varied final states
and using different approximations. In the upper panel of Fig.~\ref{fig:svtt-g},
we show the LO contribution to the $SS\to t\bar t$ channel, whilst the NLO
corrections, computed in the approximation of Eq.~\eqref{eq:svttgall}, are
included in the central panel. In the lower panel of the figure, we only show the gluon emission contributions, $SS\to t \bar t g$, computed in the
chiral limit for $m_S>300$ TeV. The (loop-induced) contributions of the $SS\to gg$ channel to the
annihilation cross section are evaluated in Fig.~\ref{fig:svg}.

Comparing the upper and central panels of Fig.~\ref{fig:svtt-g}, we
observe that QCD emissions play a significant role for $m_S>2$ TeV, as
already visible in Fig.~\ref{fig:svtt-g-ratios} (in which the exact
NLO results from Ref.~\cite{Colucci:2018qml} have been employed). In contrast,
Fig.~\ref{fig:svg} shows that annihilations into pairs of gluons are
only relevant for $m_S < m_t$ (see also section~\ref{%
  sec:relic}). Moreover, $\sigma v_{gg}$ exhibits a minimum around
$m_S\sim 280$~GeV independently of the value of $r$. This minimum is
connected to a change of sign at the level of the loop-amplitude that
always happens for $m_S \in [270,290]$~GeV (see
appendix~\ref{sec:ggloop} for an analytic expression of $\sigma
v_{gg}$). As in Fig.~\ref{fig:viab}, the yellow region around $m_S \sim
60-70$ GeV in Fig.~\ref{fig:svg} corresponds to models with a DM abundance
dominated by the resonant co-annihilation of a $TS$ system into a
top quark.

\begin{figure}
  \begin{center}
    \includegraphics[width=0.98\columnwidth]{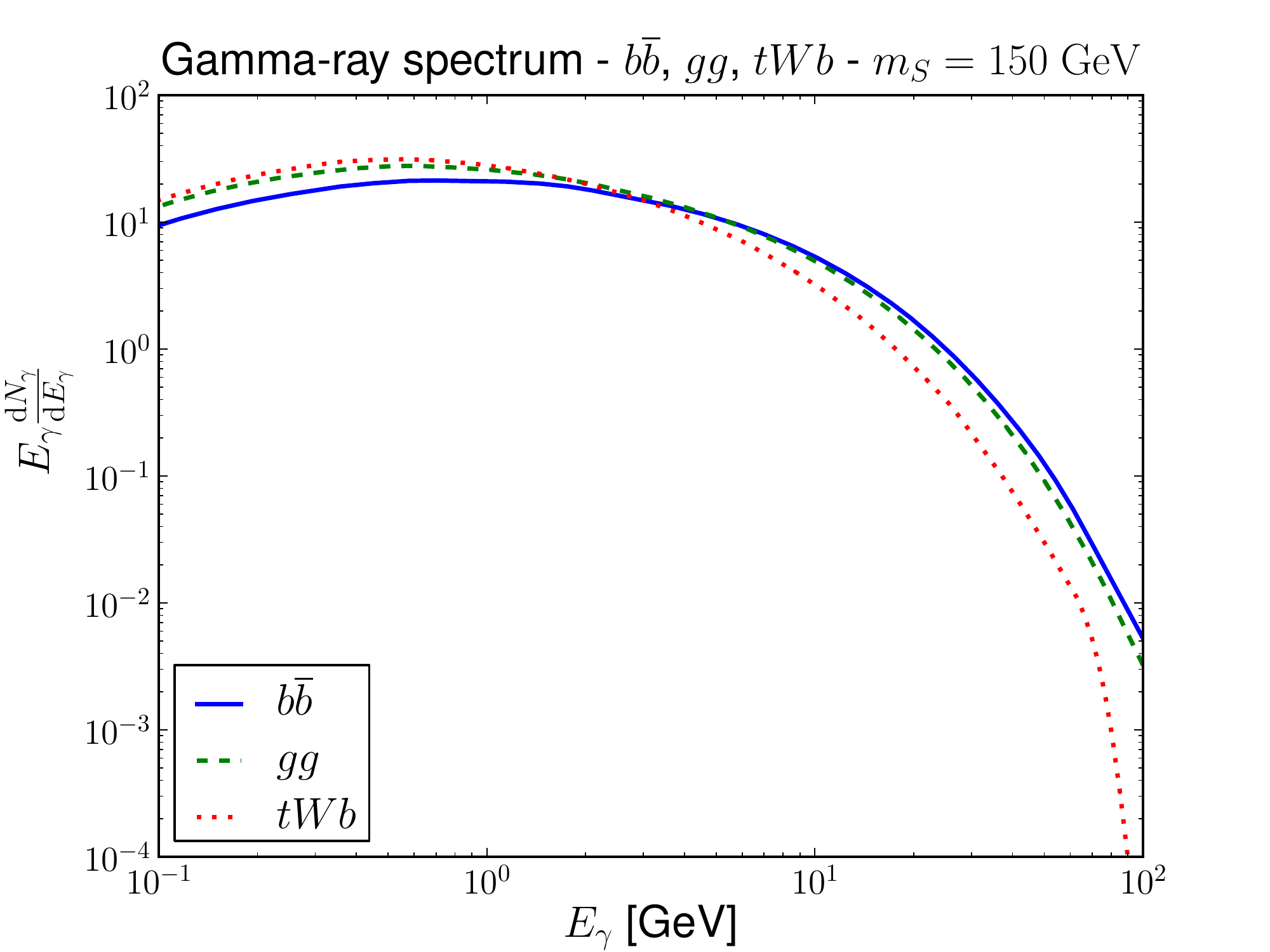}
    \includegraphics[width=0.98\columnwidth]{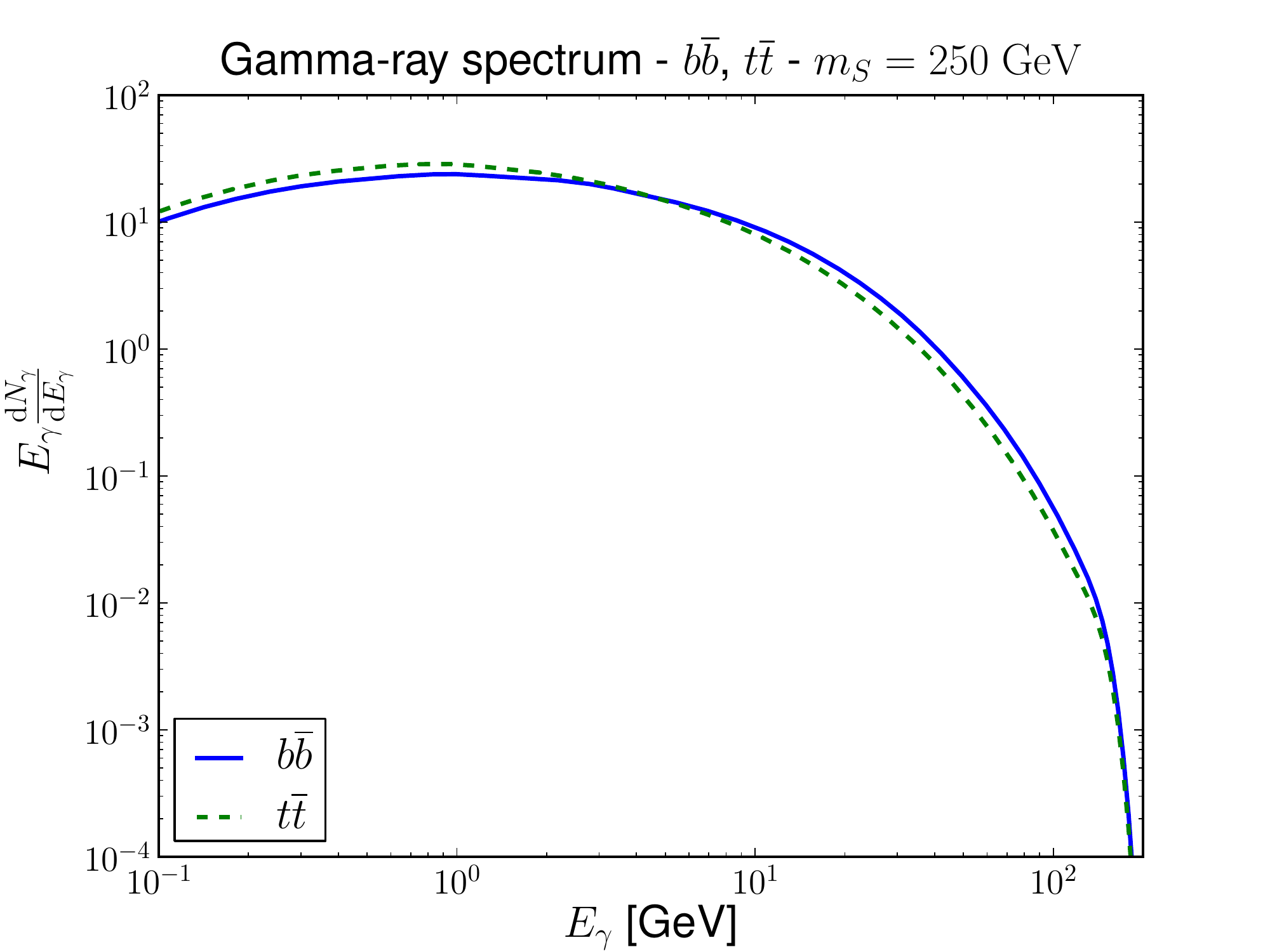}
    \caption{Gamma-ray spectra as obtained with
      {\sc Pythia}~8~\cite{Sjostrand:2014zea}  for different
      DM annihilation mechanisms. We consider a DM mass of $m_S = 150$~GeV
      (upper panel) and 250~GeV (lower panel).
  \label{fig:spectra}}
  \end{center}
\end{figure}

We superimpose to our predictions limits extracted from varied
observations. DM annihilations into top-antitop systems can be
constrained with antiproton cosmic ray data~\cite{Cuoco:2017iax}
(continuous green lines in Fig.~\ref{fig:svtt-g}). We also show
Fermi-LAT gamma-ray constraints from dwarf spheroidal analysis for
annihilations into a $b \bar b$ final state~\cite{Ackermann:2015zua}
(continuous dark green line in Fig.~\ref{fig:svg}) and the
corresponding prospects from 15 years of Fermi-LAT
running~\cite{Charles:2016pgz} (dot-dashed orange lines). Whilst the
Fermi-LAT collaboration has not published any specific limits for what
concerns the gamma-ray spectrum issued from DM annihilations into the
$t \bar t$ and $gg$ final states, both spectra are expected to show a
similar behavior as for annihilations into a $b\bar b$ system, as
illustrated in Fig.~\ref{fig:spectra} for $m_S$ below (upper panel) and above
(lower panel) the top
mass. An estimate of the limits for $t \bar t$ and $gg$ final states can
be obtained following the methodology advocated
in Ref.~\cite{Bringmann:2012vr}, using exclusion limits from DM
annihilations into $b \bar b$ pairs that are rescaled using
\begin{equation}
  \sigma v_{gg, t\bar t} = \sigma v_{b \bar b} \frac{N_\gamma^{b\bar{b}}}
    {N_\gamma^{gg,t \bar t}} \ ,
\end{equation}
where $N_\gamma^{X}$ is the number of photon expected from an $X$
final state. We have nevertheless verified that $N_\gamma^{b \bar b} <
N^{gg,t \bar t}$ by determining $N_\gamma^{X}$ using the hadronization
model of {\sc Pythia}~8~\cite{Sjostrand:2014zea}, so that the obtained
bounds can be seen as conservative.

\begin{figure}
  \begin{center}
    \includegraphics[width=0.98\columnwidth]{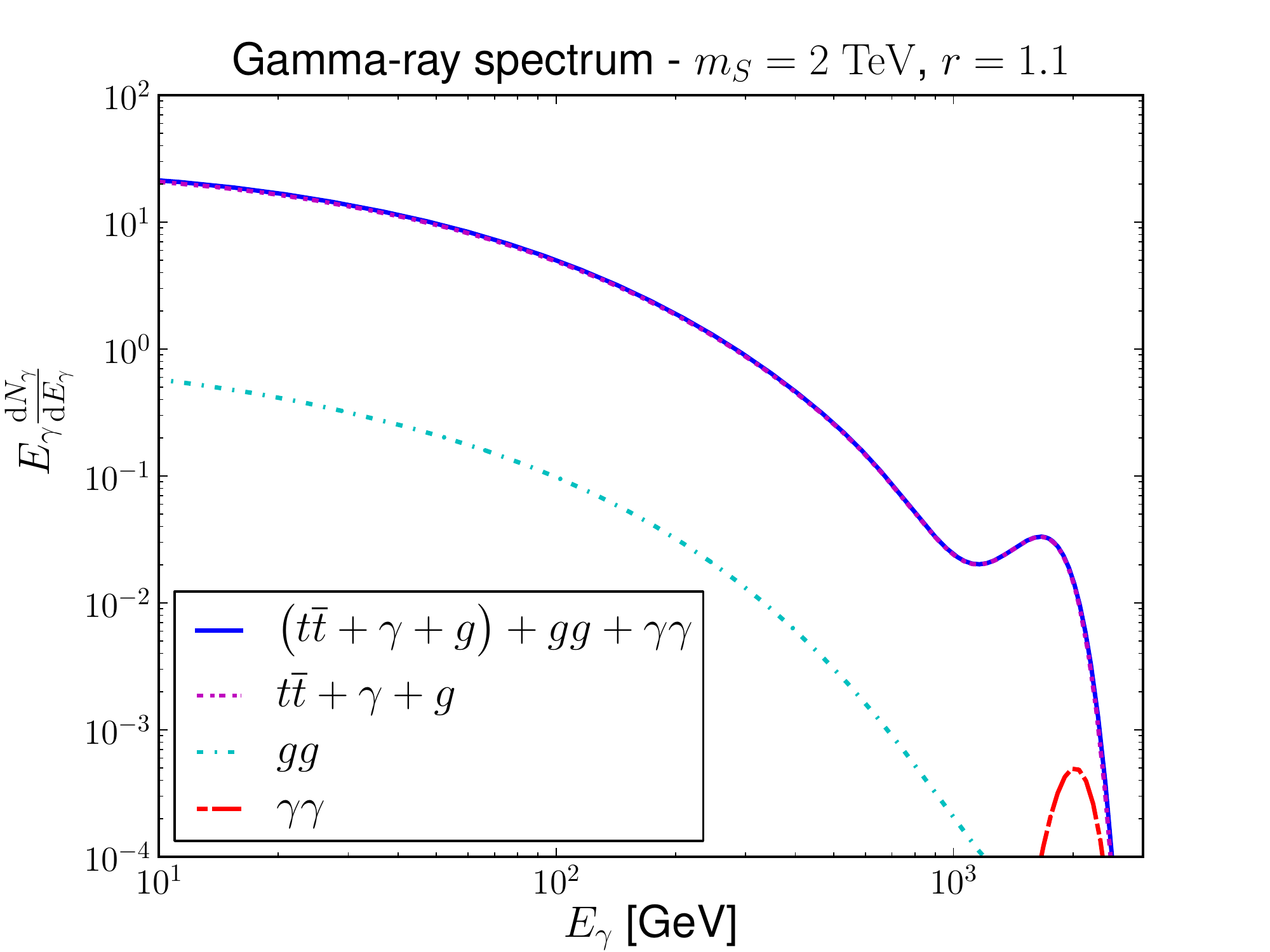}
    \includegraphics[width=0.98\columnwidth]{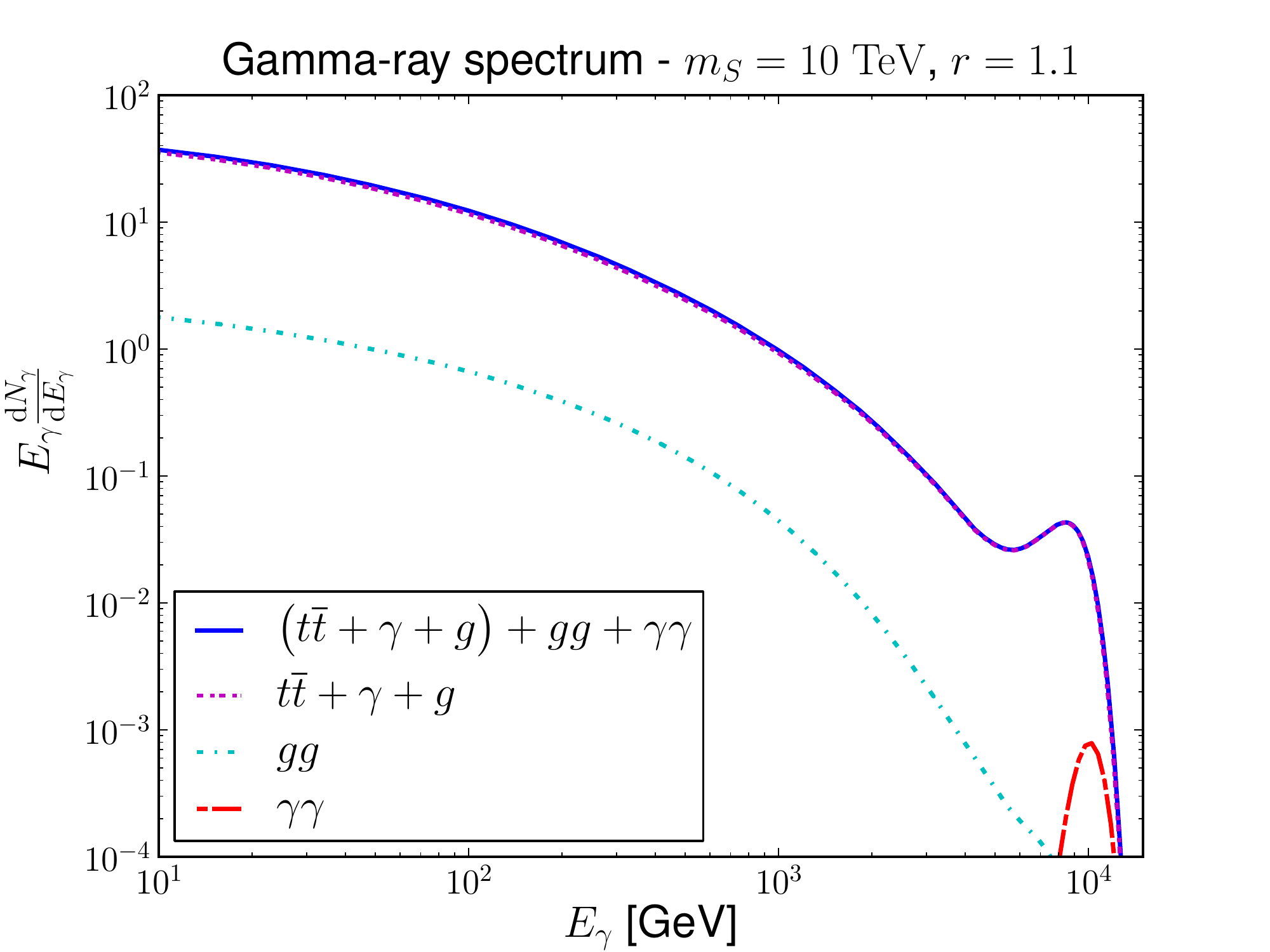}
  \caption{Gamma ray spectrum originating from the annihilation of a pair of
    $S$ particles of mass $m_S = 2$~TeV (upper panel) and 10~TeV (lower panel),
    and for a mediator mass fixed through the $r$ parameter that is set to
    $r=1.1$.
    Our predictions include (virtual and final state) gluon and photon
    emissions from a $t\bar t$ final state, as well as the direct one-loop
    contributions issued from annihilations into a pair of monochromatic photons
    and gluons.
  \label{fig:svtt-spect1}}
  \end{center}
\end{figure}

The shape of the gamma-ray spectrum could also potentially be used to
get hints on DM, as radiative corrections may give rise to specific
gamma-ray spectral imprint such as line-like features. However, these
are most of the time overwhelmed by the continuum originating from the
hadronization of the annihilation products. There are however
two regimes in which they may be potentially important, namely in the low mass range
($m_S<m_t$) where annihilations into a photon pair could be relevant,
and in the multi-TeV regime where radiative emission is crucial (as
shown on the different panels of Fig.~\ref{fig:svtt-g}).  The typical
gamma-ray spectral signature of the annihilation of a pair of very
heavy $S$ particles into $t\bar t$, $\gamma\gamma$ and $gg$ systems is
presented in Fig.~\ref{fig:svtt-spect1}, our predictions being derived
as sketched in appendix~\ref{sec:QCDcorrections}.

\noindent$\bullet$ \underline{$\mathbf{m_S \gtrsim 5}$ {\bf TeV}}. This regime is the one for which
VIB emissions play a significant role and for which the approximation of
Eq.~\eqref{eq:svttgall} holds. DM annihilations into a top-antitop system
produced in association with a photon can then be simply deduced,
\begin{equation}
{\sigma v_{t \bar t \gamma} \over \sigma v_{t \bar t g}} = {2 N_c Q^2 \alpha \over (N_c^2-1)\alpha_s}\approx 2.3\cdot 10^{-2} \ ,
\label{eq:rat-ttgamg}
\end{equation}
where $N_c=3$ denotes the number of
colors. Moreover, $\alpha$ and $\alpha_s$ stand for the electromagnetic
and strong coupling constants and we use $Z$-pole values as references,
$\alpha=1/128$ and $\alpha_s=0.112$. Although results from the
H.E.S.S. collaboration can potentially constrain the model, there is
no official VIB dedicated analysis and one must thus refer to the
independent analysis of Ref.~\cite{Ibarra:2013eda} and the recent
constraints that can be extracted from the gamma ray spectrum issued
from the galactic center~\cite{Rinchiuso:2017kfn}.  This suggests that
the annihilation cross section can be of at most $\sigma v_{t\bar
  t\gamma} \sim 10^{-27}$~cm$^3/$s for DM masses of about 10~TeV,
which can be translated as $\sigma v_{t\bar t g} \sim
10^{-25}$~cm$^3/$s. This is illustrated in the lower panel of
Fig.~\ref{fig:svtt-g} where we show the H.E.S.S.  constraints derived
in Ref.~\cite{Rinchiuso:2017kfn}, after including both the rescaling
factor of Eq.~\eqref{eq:rat-ttgamg} and a factor of $2$ accounting for
the photon multiplicity.

\noindent$\bullet$ \underline{$\mathbf{m_S < m_t}$}. In this regime, $\sigma v_{gg}$ can be as
large as about $2 \cdot 10^{-26}$~cm$^3/s$ (see Fig.~\ref{fig:svg}), and there
is a well defined prediction for annihilations into a pair of
photons~\cite{Chu:2012qy},
\begin{equation}
{\sigma v_{\gamma\gamma} \over \sigma v_{gg}} = \frac{4Q^4\alpha^2N_c^2}{\alpha_S^2\left(N_c^2-1\right)} \approx 4.3 \cdot 10^{-3} \,.
\label{eq:gamgrat}
\end{equation}
The strongest constraints on the production of
gamma-ray lines at energies around and below $m_t$ originate from the
Fermi-LAT collaboration~\cite{Ackermann:2015lka} and we indicate them
in Fig.~\ref{fig:svg} after including the rescaling factor of
Eq.~\eqref{eq:gamgrat} (gray dotted line). H.E.S.S. bounds at larger
DM masses are also indicated, following Ref.~\cite{Rinchiuso:2017kfn}
(double-dot-dashed line). In both cases, we use the limits associated
with an Einasto DM density profile.

To conclude this section, we project the DM indirect detection constraints from
the cosmic ray analysis (green region at large mass) and existing (dark green
region at low mass) and future (orange region with a dot-dashed contour)
Fermi-LAT constraints from the gamma-ray continuum from dwarf
Spheroidal galaxies in the summary of Fig.~\ref{fig:summary}.
The color code is the same as in
Figs.~\ref{fig:svtt-g} and~\ref{fig:svg}.  A substantial part of the
parameter space, for $m_S< 1$ TeV region, turns out to be constrained by
probes of the gamma-ray continuum and antiproton cosmic
rays. Moreover, for moderately heavy DM candidates, these constraints
are complementary to those originating from direct DM searches studied
in Sec.~\ref{sec:direct-detect-constr}.  As for the relic density,
annihilations into pairs of gluons are relevant for light DM ($m_S <
m_t$) whilst annihilations into top-antitop systems help to test
heavier candidates with masses ranging up to $m_S\sim 400$~GeV and
$450$~GeV when observations based on gamma rays and antiprotons are
respectively used. The major difference with the relic density
considerations is that close to the top-antitop threshold, the
non-zero DM velocity at the freeze-out time allows for DM
annihilations into a $t \bar t$ pair, which is kinematically forbidden
today. A three-body $tWb$ final state must therefore be considered
instead, which does not yield further constraints. Finally, the
predicted annihilation cross sections $\sigma v_{gg}$ and $\sigma
v_{t\bar t g}$ appear to be too small to allow us to constrain the
models using searches of specific features in the gamma-ray spectrum
(considering an Einasto DM density profile).

\section{Collider constraints}
\label{sec:lhc}

Searches for new physics have played an important role in past, current and
future physics programs at colliders. In the context of the class of scenarios
investigated in this work, in which the Standard Model is extended by a bosonic
DM candidate and a fermionic vector-like mediator, the results of many
collider analyses can be reinterpreted to constrain the model.

In our model, the extra scalar particle is rendered stable (and thus a viable
candidate for DM) by assuming a $\mathbb{Z}_2$ symmetry under
which all new states are odd and all Standard Model states are even. As a
consequence, the collider signatures of the model always involve final states
containing an even number of odd particles that each decay into Standard Model
particles and a DM state. This guarantees the presence of a large
amount of missing transverse energy as a generic model signature.

For top-philic models, the relevant signatures can be classified into two
classes, the model-independent mono-X searches that target the production of a
pair of DM particles in association with a single energetic visible
object X, and the production of a pair of top-antitop quarks
in association with missing energy.

Before going through the most recent constraints originating from LHC searches
for DM, we will account for LEP results. In electron-positron
collisions, top partners can be produced electroweakly,
\begin{equation}
  e^+ e^- \to \gamma^*, Z \to T \bar T \to t \bar t + \slashed{E}_T \ ,
\end{equation}
and yield a signature made of a pair of top-antitop quarks and missing
transverse energy $\slashed{E}_T$. Reinterpreting the results of the
LEP searches for the supersymmetric partner of the top quark,
vector-like (top) partners are essentially excluded if their mass
satisfies $m_T\lesssim 100$~GeV~\cite{Abbiendi:2002mp}. This excludes
the lower left corner of the viable parameter space of the summary
of Fig.~\ref{fig:summary} (magenta region) corresponding to DM
masses of typically $m_S<78$ GeV.

\begin{figure}
  \centering
  \includegraphics[width=0.32\columnwidth]{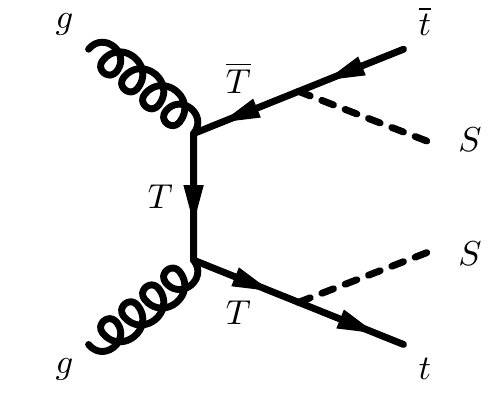}
  \includegraphics[width=0.32\columnwidth]{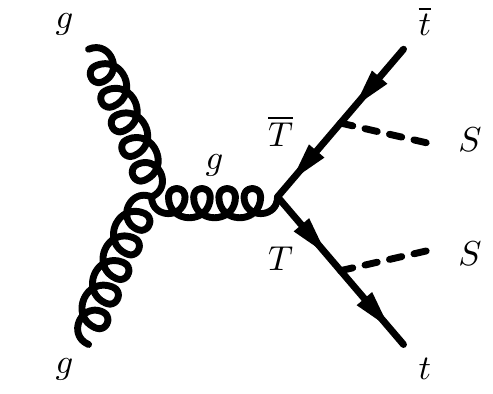}
  \includegraphics[width=0.32\columnwidth]{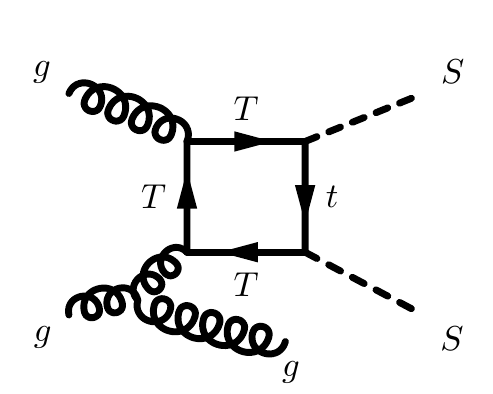}
  \caption{Representative Feynman diagrams corresponding to the collider
    signatures of the model.
    We consider the pair production of vector-like quarks that
    then decay each into a DM state and a top quark (leftmost and
    central diagrams) and loop-induced monojet production (rightmost diagram).}
\label{fig:collider_graphs}
\end{figure}

At the LHC, pairs of mediators can be copiously produced by virtue of the strong
interaction. Top-antitop production in association with missing energy consists
of the corresponding signature, as each mediator then decays, with a 100\%
branching fraction, into a system comprised of a top quark and a DM
particle,
\begin{equation}
  p p \to T \bar T \to t S \ \bar t S\ .
\label{eq:pp2TT}\end{equation}
Contributions to this process are illustrated by the first two Feynman diagrams
of Fig.~\ref{fig:collider_graphs}.
Such a top-antitop plus missing energy signature has been widely studied by
both the ATLAS and CMS collaborations, in particular in Run~2 searches
for the superpartners of the top quark (assuming a decay into a top
quarks and missing energy carried by a neutralino)~\cite{Aaboud:2017aeu,%
Aaboud:2017ayj,Aaboud:2017wqg,Aaboud:2017nfd,Aaboud:2017dmy,Sirunyan:2017kqq,%
Sirunyan:2017wif,Sirunyan:2017xse,Sirunyan:2017leh} and in dedicated DM
searches~\cite{Sirunyan:2017xgm}.

Additionally, the model can also be probed through classical DM
searches using mono-X probes. Amongst all mono-X searches, we
focus on the monojet one given the relative magnitude of the strong coupling
with respect to the strength of the electroweak interactions. In this case, the
considered signature exhibits the presence of a hard QCD jet recoiling against
a large quantity of missing energy carried away by a pair of DM
particles. Such a process,
\begin{equation}
  p p \to S S j\ ,
\label{eq:monoj}\end{equation}
is loop-induced in our model, as illustrated by the last Feynman diagram of
Fig.~\ref{fig:collider_graphs}. Although early monojet analyses were vetoing
events featuring any extra hadronic activity through additional hard jets, it
has been demonstrated that the latter could consist in useful handles to get a
better sensitivity to the signal~\cite{Buchmueller:2015eea}. For this reason,
recent ATLAS and CMS monojet analyses now include several signal regions in
which more than one hard jet is allowed~\cite{Aaboud:2016tnv,Aaboud:2017phn,%
Sirunyan:2017hci,Sirunyan:2017jix}.

\subsection{Simulation details}
\label{sec:simu}
In order to reinterpret relevant results of the LHC in the context of the
considered top-philic DM scenario and to determine their impact,
we have implemented the Lagrangian of
Eq.~\eqref{eq:lag} into the {\sc FeynRules} program~\cite{Alloul:2013bka}. With
the help of a joint usage of the NLOCT~\cite{Degrande:2014vpa} and
{\sc FeynArts}~\cite{Hahn:2000kx} packages, we have analytically evaluated the
ultraviolet and so-called $R_2$ counterterms required for numerical one-loop computations
in four dimensions. The information has been exported under the
form of an NLO UFO model~\cite{Degrande:2011ua} containing, in addition to
the tree-level model information, the $R_2$ and NLO counterterms.

We rely on the
{\sc MadGraph~5}\_aMC@NLO~\cite{Alwall:2014hca} platform for the generation of
hard-scattering events, at the NLO accuracy in QCD for the vector-like quark
pair production process of Eq.~\eqref{eq:pp2TT} and at the LO accuracy
for the loop-induced monojet process of Eq.~\eqref{eq:monoj}. In our simulation
chain, we respectively convolute the LO and NLO matrix elements with the LO and
NLO sets of NNPDF~3.0 parton distribution functions~\cite{Ball:2014uwa}, that we
access through the LHAPDF~6 library~\cite{Buckley:2014ana}. Moreover, the
unphysical scales are always set to half the sum of the transverse mass of all
final-state particles.

The decay of  the heavy $T$ quark into DM and a top quark,
\begin{equation}
  T \to t S \ ,
\end{equation}
is factorized from the production processes and is handled with the
\ms~\cite{Artoisenet:2012st} and \mw~\cite{Alwall:2014bza} programs, together
with those of all Standard Model heavy particles. For each considered new
physics setup, we have consequently checked that the narrow-width approximation
could be safely and consistently used, which is guaranteed by the fact that the
mediator decay width satisfies $\Gamma_T/m_T<0.2$.

The resulting
partonic events are matched with parton showers by relying on the
{\sc Pythia}~8 code~\cite{Sjostrand:2014zea} and the MC@NLO
prescription~\cite{Frixione:2002ik}. Whilst hadronization is also taken care of
by {\sc Pythia}, we simulate the response of the ATLAS and CMS detectors
by means of the \del~3 program~\cite{deFavereau:2013fsa} that
internally relies on the anti-$k_T$ jet algorithm~\cite{Cacciari:2008gp} as
implemented in the \fj~software~\cite{Cacciari:2011ma} for object
reconstruction. For each of the analyses that we have recast, the \del\
configuration has been tuned to match the detector setup described in the
experimental documentation. We have used the \ma\
framework~\cite{Conte:2012fm,Conte:2014zja,Dumont:2014tja} to calculate the
signal efficiencies for the different considered search strategies and to derive
95\% confidence level (CL) exclusions with the CL$_s$ method~\cite{Read:2002hq}.

\subsection{Reinterpreted LHC analyses}
\label{se:reinter}
In order to assess the reach of LHC searches for DM in top-antitop
quark production in association with missing energy ($pp\to t\bar t +
\slashed{E}_T$), we reinterpret a CMS
analysis of collision events featuring a pair of leptons of opposite
electric charge~\cite{Sirunyan:2017leh}. While other final states in the single
lepton and fully hadronic decay mode of the top-antitop pair are
relevant as well, all these LHC searches are so far found to
yield similar bounds. For this reason, we have chosen to
focus to a single of those channels, namely the cleaner dileptonic decay mode of
the top-antitop pair.

The CMS-SUS-17-001 analysis of Ref.~\cite{Sirunyan:2017leh} focuses on the
analysis of 35.9~fb$^{-1}$ of LHC collisions featuring the presence of a system
of two isolated
leptons of opposite electric charges which is compatible neither with a low-mass
hadronic resonance nor with a $Z$ boson. The presence of
at least two hard jets is required, at least one of them being $b$-tagged, as
well as a large amount of missing transverse energy. The latter is required to
possess a large significance and to be well separated from the two leading jets.
After this preselection, the analysis defines three aggregated signal regions
depending of the value of the missing energy and of the transverse
mass $m_{T2}$~\cite{Lester:1999tx,Cheng:2008hk} reconstructed from the two
leptons and the missing momentum.

In addition, we include in our investigations the CMS-SUS-16-052 analysis which
is dedicated to probing the more compressed regions of the parameter space with
35.9~fb$^{-1}$ of LHC collisions~\cite{CMS:2017odo}. In this analysis, it is
assumed that the top partner cannot decay on-shell into a top quark plus missing
energy system, so that the search strategy is optimized for top partners
decaying into systems made of three mostly soft fermions (including $b$-quarks)
and missing energy via an off-shell top quark. Event selection requires the
presence of one hard
initial-state-radiation jet and of at most a second jet well separated from the
first one. Moreover, one asks for a single identified lepton, a large amount of
missing energy and an important hadronic activity. The threshold values that are
imposed and the detailed properties of the lepton, the missing energy and the
hadronic activity allow to define two classes of three signal regions targeting
varied new physics configurations.

We have also confronted the process of Eq.~\eqref{eq:pp2TT} to the LHC
Run~1 results, and in particular to the null results of the 8~TeV search
labeled CMS-B2G-14-004~\cite{Khachatryan:2015nua,Arina:2016cqj,%
inspire-cms-b2g-14-004}. This search focuses on
singly-leptonic final states containing at least three jets (including at least
one $b$-tagged jet) and a large amount of missing energy well separated from the
jets. The event selection moreover constrains the transverse mass of the system
comprised of the lepton and of the missing transverse momentum, as well as the
$m_{T2}^W$ transverse variable~\cite{Bai:2012gs}.

For the reinterpretation of the LHC search results for mono-X DM
signals, we have considered two ATLAS analyses targeting
a monojet-like topology, {\it i.e.} at least one very hard jet recoiling
against some missing momentum and a subleading jet activity. Although those
analyses~\cite{Aaboud:2016tnv,Aaboud:2016zdn} focus on a small integrated
luminosity of LHC collisions (3.2~fb$^{-1}$), they are already limited by the
systematics so that the constraints derived from early Run~2 data
are not expected to get more severe in the future~\cite{Banerjee:2017wxi}.
In the ATLAS-EXOT-2015-03 analysis~\cite{Aaboud:2016tnv,%
inspire-atlas-exot-2015-03}, the target consists in
a monojet-like topology where the subleading jet activity is rather limited, the
event selection being allowed to contain only up to three additional jets.
Seven inclusive and seven exclusive signal regions are defined, the
differences between them being related to various requirements on the missing
energy. In contrast, the ATLAS-SUSY-2015-06 analysis~\cite{Aaboud:2016zdn,%
inspire-atlas-susy-2015-06} allows both for a small and larger subleading jet
activity, the event selection being dedicated to final states containing
two to six jets. Seven signal regions are defined, depending on the number and
on the kinematic properties of the jets and on the missing momentum.

All the above analyses are implemented and validated in the \ma\ framework, and
have thus been straightforwardly and automatically used within the simulation
chain depicted in section~\ref{sec:simu}. We consider a new physics signal
including contributions from both processes of Eq.~\eqref{eq:pp2TT} and
Eq.~\eqref{eq:monoj}, although vector-like quark pair production largely
dominates for perturbative $\tilde{y}_t$ values.

\subsection{Collider constraints}
\begin{figure}
  \begin{center}
       \includegraphics[width=8.5cm]{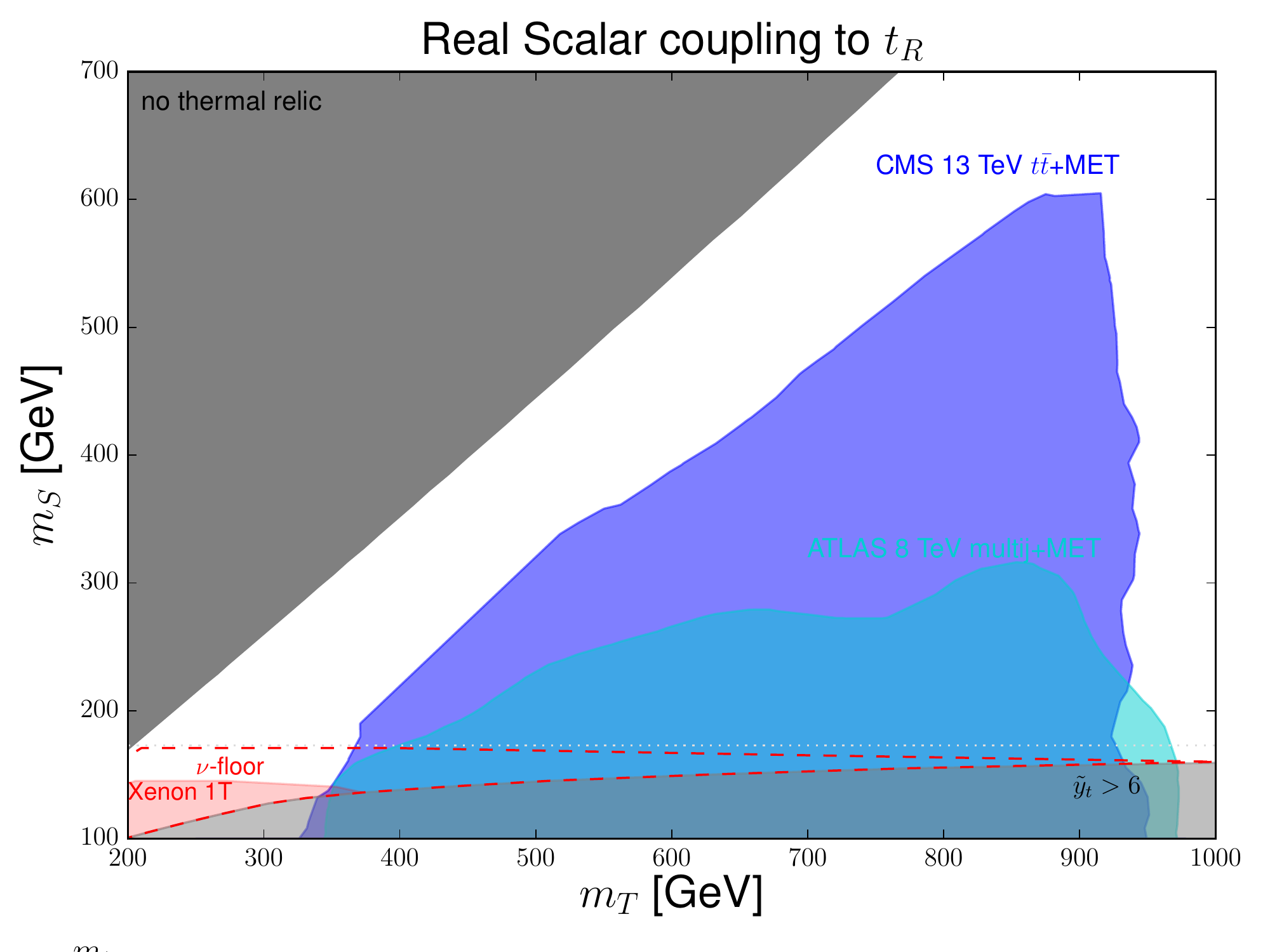}
  \caption{Collider constraints on our top-philic DM model expressed, together
    with the relic density and DM direct detection bounds, in the $m_S, m_T)$
    mass plane. \label{fig:coll}}
  \end{center}
\end{figure}

In Fig.~\ref{fig:coll}, we report our findings in the $(m_T, m_S)$ mass plane. 
As the vector-like-quark production process of Eq.~\eqref{eq:pp2TT}
dominates regardless of the actual value of the $\tilde{y}_t$ coupling, the
latter is irrelevant for what concern constraints stemming from the LHC. This is
induced by the fact that the vector-like mediator only couples to the top
quark, which contrasts with scenarios in which couplings to lighter quarks
exist. Those interactions with the first and second generation SM quarks indeed
yield extra contributions featuring a direct dependence on the Yukawa
couplings~\cite{Giacchino:2015hvk}. Coming back to the considered top-philic
scenario, all results can thus be represented in the $(m_S, m_S/m_T-1)$
two-dimensional plane. In the
figure, we superimpose to the cosmology considerations discussed in the previous
sections (namely the relic density and direct detection bounds, the indirect
bounds being not reproduced, so as to avoid cluttering of the figure) the
constraints
that can be obtained by reinterpreting the results of the LHC searches for new
physics discussed in section~\ref{se:reinter}. The white region corresponds to
configurations for which the experimentally-observed relic abundance is
reproduced and which are not excluded by current cosmological data. In the
(excluded) light gray region, a correct abundance would imply going beyond the
perturbative regime, whilst in the dark gray region, the dark matter particle
$S$ is unstable as not the lightest $\mathbb{Z}_2$-odd particle.

For each new physics configuration and each signal region of each considered
analysis, we evaluate the number of signal events $s$ surviving the selection
with {\sc MadAnalysis}~5 and extract a CL$_s$ exclusion from the
observed number of events $n_{\rm data}$ populating the region and
the expected number of background events $\hat b\pm \Delta b$. To this aim, we
undertake 100.000 toy experiments in which we generate the actual number of
background events $b$ by assuming that the corresponding distribution is a
Gaussian of mean $\hat b$ and width $\Delta b$. We then consider two Poisson
distributions of parameters $b$ and $b+s$ to evaluate the $p$-values of the
signal-plus-background and background-only hypotheses, knowing that
$n_{\rm data}$ events have been observed. From these $p$-values, we
derive the associated CL$_s$ value.

The colored regions shown in Fig.~\ref{fig:coll} are excluded at the 95\% CL by
at least one signal region of the considered analyses. The dark blue region
corresponds to what we obtain with the reinterpretation of the results of the
two CMS searches for DM in the top-antitop plus missing energy channel,
namely CMS-SUS-17-001 and CMS-SUS-16-052. Whilst our results only focus on Run~2
data, we have verified that the obtained limits are compatible with the less
stringent Run~1 constraints derived from the results of the CMS-B2G-14-004
analysis. The light blue area depicted on the figure corresponds to bounds that
can be extracted from the reinterpretation of the results of the
ATLAS-EXOT-2015-03 and ATLAS-SUSY-2015-06 searches for new physics in the
multijet plus missing energy channel.

We have found that mediator masses ranging up to 1~TeV are excluded, provided
that the DM mass is light enough for having enough phase space to
guarantee the decay of the mediator into a DM particle and a top quark
in a far-from-threshold regime. Whilst generic multijet plus missing energy
searches are quite sensitive when the DM mass is small, they quickly
lose any sensitivity for larger $m_S$ values. This stems from the monojet-like
selection of the considered analyses, that can only be satisfied if enough
phase space is available for the $T$ decay process.

As soon as the $T\to t S$ decay channel is closed, the $T$ quark becomes
long-lived enough to hadronize before decaying and it could  potentially travel
on macroscopic distances in the detector. Whilst the unknown modeling of
vector-like quark hadronization would introduce uncontrolled uncertainties on
the predictions, none of the currently available computer tools allows for
a proper handling of long-lived colored particles. Moreover, all considered LHC
analyses have been designed for being sensitive to promptly-decaying new-physics
states, and are thus expected to lose sensitivity when new physics particles are
long-lived. For this reason, we restrict ourselves to provide LHC constraints in
the region of the parameter space where the $T$ quark can promptly decay into a
top quark and a DM particle.


\section{Summary}
\label{sec:summary}

\begin{figure}
  \centering
  \includegraphics[width=.98\columnwidth]{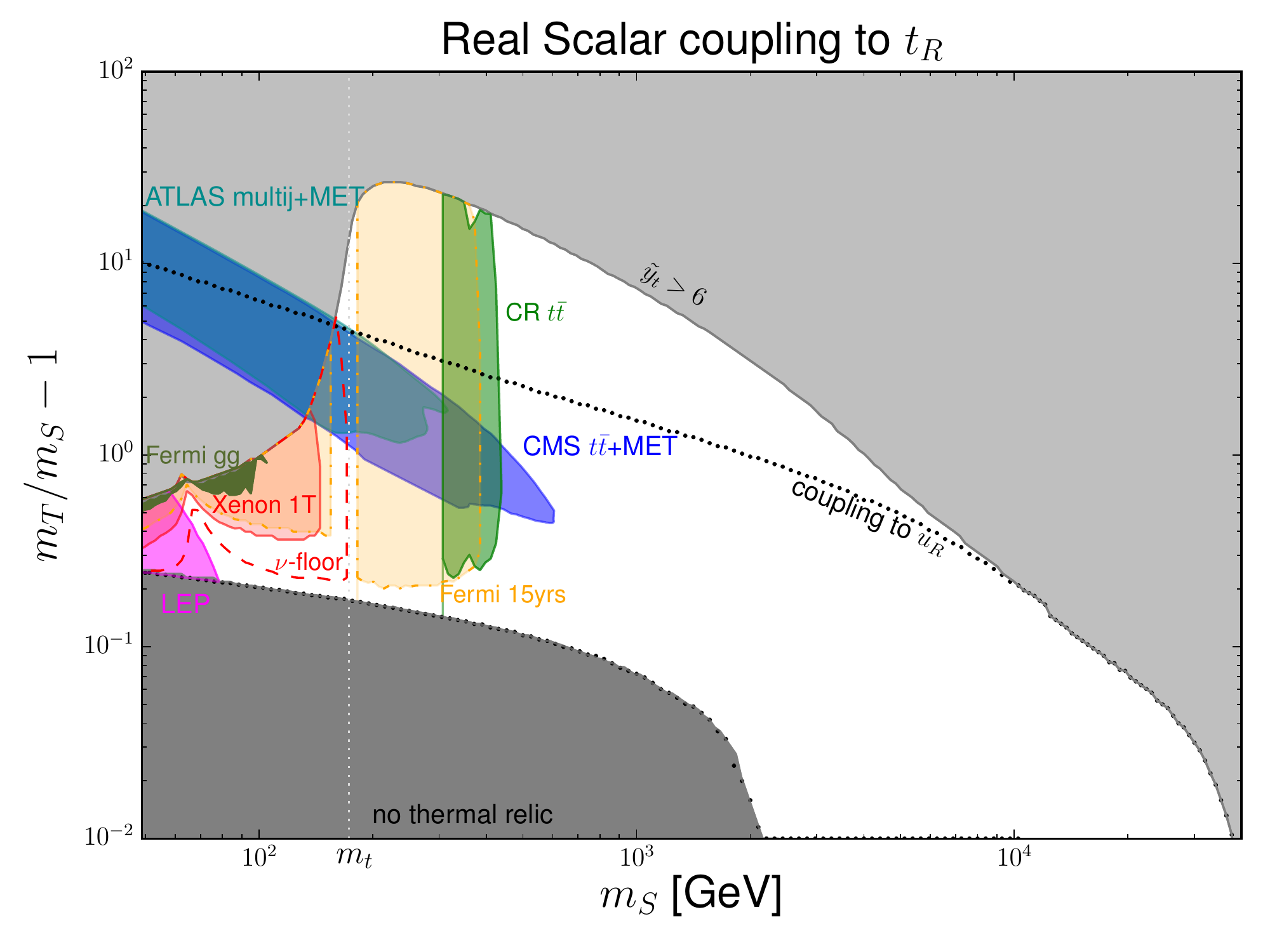}
  \caption{Phenomenologically-viable region of our model parameter space,
    presented in the $(m_S, r-1)$ plane, on which we project constraints from
    DM direct and indirect detection and collider experiments. The grey regions
    correspond to regions for which the relic density cannot be
    accommodated. {\it Direct DM searches:} The red region is excluded
    by the Xenon 1T~\cite{Aprile:2017iyp} experiment at the 90\%
    confidence level while the region delimited by the red dashed line
    is in principle testable by DM direct detection searches as lying
    above the neutrino floor~\cite{Billard:2013qya}. {\it Indirect DM
      searches:} the dark green (at low mass) and light green
    (at large mass) regions are excluded by Fermi-LAT gamma-rays
    constraints~\cite{Ackermann:2015zua} and by the CR analysis
    of Ref.~\cite{Cuoco:2017iax}. The orange region delimited by a dot-dashed
    line is the projected sensitivity of Fermi-LAT after 15
    years of exposure~\cite{Charles:2016pgz}. {\it Collider searches:}
    constraints on top partner production at LEP~\cite{Abbiendi:2002mp} and the
    LHC~\cite{Sirunyan:2017leh,CMS:2017odo} are respectively shown by the
    magenta and light blue regions, whilst
    mono-X bounds~\cite{Aaboud:2016tnv,Aaboud:2016zdn} are indicated by the dark
    blue region.}
\label{fig:summary}
\end{figure}

The WIMP paradigm is being tested by various experiments, both in
astrophysics, cosmology and at colliders. At the same time, there is a
significant interest on top-philic new physics, as the top quark is
widely considered, due to its large mass, as a perfect laboratory for
the study of the electroweak symmetry breaking mechanism. In this
work, we have extensively investigated a simple DM scenario that
brings naturally these topics together. It is based on a real scalar
particle coupled to the top quark through a Yukawa coupling with a
heavy vector-like quark. As in the SM sector, the top quark has the
largest coupling to the Higgs boson, it is at least conceivable that
it also features the largest coupling to a new dark sector. The model
rests only on very few parameters (one coupling strength and two
masses), so that it provides a good starting point to compare the
impact of different experimental results from varied origins. In the
present case, we focus on DM direct and indirect detection searches,
as well as on collider probes.  We have studied the constraints on
the DM model, paying special attention to the potential impact of the
QCD radiative corrections on all the considered bounds ({\it i.e.},
the DM relic abundance, the DM direct and indirect searches and the
collider searches). In this way, our study complements and extents similar
earlier works based on Majorana DM candidates~\cite{Ibarra:2015nca,%
Bringmann:2015cpa,Garny:2018icg}.

Our analysis reveals that, although there is a complementarity between
the different searches, only a small fraction of the viable parameter
space of this very simple DM scenario is tested by the current
experiments. This is illustrated in Fig.~\ref{fig:summary} which
summarizes our results and complements the information provided in
Fig.~\ref{fig:coll}. On the long term, the most fruitful strategy
to further test such a DM scenario would be to increase the energy reach at
colliders. 

\appendix

\section{Sommerfeld corrections to the relic abundance}
\label{sec:somm-corr}

DM annihilation can proceed directly through the
\begin{equation}
  SS \rightarrow t\bar t^{(*)},\ \ gg \ \  \text{and}\ \ t\bar t g
\end{equation}
channels. If the vector-like fermionic mediator is not too heavy relatively to
the DM particle, it may still be abundant at the time of freeze-out and hence
either annihilates or co-annihilates~\cite{Griest:1990kh},
\begin{equation}\bsp
  & S T \to g \,t\ ,\qquad S \bar T \to g \,\bar t\ ,\\ & T \bar T \to
  gg\ \ \text{or}\ \ q\bar q \ ,\\ & T T \rightarrow t\, t\ ,\qquad \bar
  T \, \bar T \rightarrow \bar t\, \bar t \ , \esp\end{equation} with the
latter processes involving $t$-channel DM exchanges. The $T \bar T$,
$T T$ and $\bar T \bar T$ channels are impacted by either attractive
or repulsive Sommerfeld effects through gluonic
exchanges~\cite{deSimone:2014pda}.  In order to account for these
effects, we have followed the procedure depicted in
Ref.~\cite{Giacchino:2015hvk} with the only difference that we have
taken explicitly into account the top-quark mass effects on the
annihilation cross sections. In addition, we have verified, through
various approaches that our treatment is correct in the $s$-wave
approximation. Our treatment agrees with the results of
Ref.~\cite{ElHedri:2017nny} in the $s$-wave approximation. Going beyond this
approximation is however known to lead to corrections to the relic density of
less than 1\%, as shown {\it e.g.}, in supersymmetry~\cite{Pierce:2017suq}.

On general grounds, we should also take into account the possible formation of
$T \bar T, T T$ and $\bar T \bar T$ bound states, which would imply a
modification of the Sommerfeld corrections. It has however been concluded, using
a setup similar to ours, that bound states have only a moderate impact on the
DM relic density~\cite{Mitridate:2017izz,Biondini:2018pwp}. As the Sommerfeld corrections affect
the relic abundance by less than at most 15\%, we ignore bound state formation
from the present calculations.

\section{$SS \rightarrow g g$ cross section}
\label{sec:ggloop}
Analytical expressions for the $SS\rightarrow gg (\gamma \gamma)$ annihilation
cross section have been given in Refs.~\cite{Ibarra:2014qma,Giacchino:2014moa}
in terms of one-loop three-points functions. Correcting a typo in
Ref.~\cite{Giacchino:2014moa}, the cross section $\sigma v$ for DM annihilation into a
gamma or gluon pair is given by
\begin{equation}
  \sigma v=\frac{2 \tilde{y}_t^4}{64\pi^3m_S^2} \left\{
    \begin{array}{ll}
      Q^4\alpha^2N_c^2 \left| {\cal M}\right|^2\ \ \text{ for photons,} \\
      \frac{\alpha_S^2\left(N_c^2-1\right)}{4}\left| {\cal M}\right|^2\ \ \text{for gluons}.
    \end{array}
\right.
\end{equation}
In this expression, the one-loop amplitude reads
\begin{eqnarray}
  &&{\cal M} = \, 2\, +\Big \{\nonumber\\
  &&\ \frac{1\!-\!r^2\!-\!z^2}{1\!+\!r^2\!-\!z^2}\frac{2z^2}{r^2\!-\!z^2}
     C_0(-m_S^2,m_S^2,0;z^2m_S^2,r^2m_S^2,z^2m_S^2)\nonumber \\
  &&\ +\frac{4z^2\left(1-z^2\right)}{1+r^2-z^2}C_0(4m_S^2,0,0;z^2m_S^2,z^2m_S^2,^2zm_S^2)\nonumber\\
  &&\ +z\leftrightarrow r\Big \},
\end{eqnarray}
where $z=\frac{m_t}{m_S}$ and
\begin{equation}\bsp
  &C_0(p_1^2,(p_1-p_2)^2,p_2^2;m_1^2,m_2^2,m_3^2)=\\
 &\ \int\frac{\mathrm{d}^4l}{i\pi^2}\frac{1}{l^2-m_1^2}\frac{1}{\left(l+p_1\right)^2-m_2^2}\frac{1}{\left(l+p_2\right)^2-m_3^2}
\esp\end{equation}
The scalar three-point functions are written~\cite{Bergstrom:1997fh}
\begin{eqnarray}
  &&C_0(4m_S^2,0,0;z^2m_S^2,z^2m_S^2,z^2m_S^2)= \frac{1}{4}I_1(1,z^2)\ ,\\
  &&C_0(-m_S^2,m_S^2,0;z^2m_S^2,r^2m_S^2,z^2m_S^2)=
    \!-\! \frac{1}{2}I_2\left(\frac{1}{z^2},\frac{r^2}{z^2}\right) \ ,
\nonumber\end{eqnarray}
with
\begin{equation}
  I_1(1,z^2) = \left\{\begin{array}{l}
  \frac{1}{2}\left[ \left(\log \frac{1+\sqrt{1-z^2}}{1-\sqrt{1-z^2}}\right)^2-\pi^2\right]\text{ if z}<\text{1}\\
  -2\left(\arctan \frac{1}{\sqrt{z^2-1}}\right)^2 \text{ if z}>\text{1}
\end{array}\right.
\end{equation}
and
\begin{equation}\bsp
  & I_2\left(\frac{1}{z^2},\frac{r^2}{z^2}\right)=
   \text{Li}_2\left(\frac{1-r^2+z^2-\sqrt{s_1}}{2z^2}\right)\\
  & \ +\text{Li}_2\left(\frac{1-r^2+z^2+\sqrt{s_2}}{2z^2}\right)\\
  & \ -\text{Li}_2\left(\frac{1-r^2+z^2-\sqrt{s_3}}{2z^2}\right)\\
  & \ -\text{Li}_2\left(\frac{-1-r^2+z^2\sqrt{s_4}}{2z^2}\right)\ .
\esp\end{equation}
In those expressions, we have introduced the variables
\begin{equation}\bsp
  s_1=&\ \left(1+r^2-z^2\right)^2-4r^2\ , \\
  s_2= & \ \left(1-r^2-z^2\right)^2-4r^2z^2\ , \\
  s_3 =&\ s_4 = \left(1+r^2+z^2\right)^2-4r^2z^2\ .
\esp\end{equation}

\section{On QCD corrections to the $SS \rightarrow t\bar t
g (\gamma)$} \label{sec:QCDcorrections}

In this section, we comment on our derivation of the differential cross section
for the $SS \rightarrow t\bar t g (\gamma)$ process at the partonic level and on
the methods that have been used to cope with the hadronization of the colored
final state particles.

At ${\cal O}(\alpha_s)$, the analytical expression for
the $SS \rightarrow t\bar t g$ amplitude includes contributions of gluon
emission by the final state quarks and the intermediate particle $T$. When
relevant, final-state radiation typically gives rise to double Sudakov
logarithms associated with soft and collinear divergences, which must be
consistently taken into account, in particular in regimes where they are large.

In our model, VIB emission is finite and moreover only relevant for the
radiation of highly-energetic gluons. When VIB contributions are negligible
({\it i.e.}, for sufficiently large $m_T/m_S$ ratios), we have simply discarded
them and used {\sc Pythia}~8~\cite{Sjostrand:2014zea} to handle both final-state
radiation and hadronization. This effectively resums the large logarithms via
the use of an appropriate Sukadov form factor. When VIB contributions are
relevant, the corresponding three-body hard-scattering process has been
explicitly evaluated with {\sc CalcHEP}~\cite{Belyaev:2012qa}, and we have made
use of {\sc Pythia}~8 to simulate the subsequent hadronization. Finally,
for low energies, we have restricted our computation to the two-body  $SS \to
t \bar t$ process and relied on {\sc Pythia}~8 for the simulation of
final-state radiation and hadronization. The matching of the separate
contributions to the gamma-ray spectrum has been achieved by implementing an
explicit cutoff on the gluon energy at the partonic level. More details can be
found in Ref.~\cite{Colucci:2018qml}, which is similar in spirit but differs in details
from the prescription proposed in Ref.~\cite{Bringmann:2015cpa}.

\medskip

\acknowledgments

We thank J. Heisig and M. Korsmeier for enlightening discussions and P.~Wu for
useful comments on our work. This study has been partly supported by French
state funds managed by
the Agence Nationale de la Recherche (ANR) in the context of the LABEX
ILP (ANR-11-IDEX-0004-02, ANR-10-LABX-63), by the FRIA, the FNRS, the
Strategic Research Program {\it High Energy Physics} and the Research
Council of the Vrije Universiteit Brussel, the IISN convention 4.4503.15 and by
the Excellence of Science (EoS) convention 30820817.

\bibliographystyle{JHEP}
\bibliography{biblio}

\end{document}